\documentclass[%
 reprint,
superscriptaddress,
 amsmath,amssymb, 
 aip,
onecolumn,
]{revtex4-2}

\usepackage{graphicx}
\usepackage{dcolumn}
\usepackage{bm}
\usepackage{hyperref}
\usepackage{xcolor}
\usepackage{subfigure} 
\usepackage{setspace}
\usepackage{ulem}
\usepackage{lineno}
\newcommand{\etal}{{\it et al. }}
\newcommand{\eg}{{\it e.g. }}
\newcommand{\ie}{{\it i.e. }}
\begin{document}

\title{Detection of magnetohydrodynamic waves by using machine learning}

\author{Fang Chen}
\author{Ravi Samtaney}%
 \email[Corresponding mail: ]{ravi.samtaney@kaust.edu.sa} 
\affiliation{%
Mechanical Engineering, Physical Science and Engineering Division, King Abdullah University of Science and Technology, Thuwal 23955-6900, Saudi Arabia 
}%
\date{\today}
\begin{abstract}
Nonlinear wave interactions, such as shock refraction at an inclined density interface, in magnetohydrodynamic (MHD) lead to a plethora of wave patterns with myriad wave types. Identification of different types of MHD waves is an important and challenging task in such complex wave patterns. Moreover, owing to the multiplicity of solutions and their admissibility for different systems, especially for intermediate-type MHD shock waves,  the identification of MHD wave types is complicated if one solely relies on the Rankine-Hugoniot jump conditions. 
MHD wave detection is further exacerbated by unphysical smearing of discontinuous shock waves in numerical simulations.   
We present two MHD wave detection methods based on a convolutional neural network (CNN) which enables the classification of waves and identification of their locations. 
The first method separates the output into a regression (location prediction) and a classification problem assuming the number of waves for each training data is fixed. 
In the second method, the number of waves is not specified {\it a priori} and the algorithm, using only regression, predicts the waves' locations and classifies its types. 
We use one-dimensional (1D) input data (density, velocity and magnetic fields) to train the two models that successfully reproduce a complex two-dimensional (2D) MHD shock refraction structure.
The first fixed output model efficiently provides high precision and recall, the accuracy of total neural network achieved is up to $0.99$, and the classification accuracy of some waves approaches unity. 
The second detection model has a relatively lower performance, with more sensitivity to the setting of parameters, such as the number of grid cells $N_{grid}$ and the thresholds of confidence score and class probability, etc.
The accuracy of the detection model attains more than $0.9$ and a high $F_1$ score of more than $0.95$.   
The proposed two methods demonstrate very strong potential to be applied for MHD wave detection in some complex wave structures and interactions.

\end{abstract}

\maketitle
\normalem
\section{\label{sec:1} Introduction}
Shock waves are a common feature in high-speed compressible flows, wherein the fluid state (density, pressure etc.) changes drastically over a thin region whose thickness is proportional to the mean free path of the gas. 
Because of the importance of shock waves in several applications, such as aircraft aerodynamics, astrophysical phenomena~\cite{Lebedev2019} and inertial confinement fusion (ICF)~\cite{Lindl2004}, etc., the detection of shock waves is an important and a  challenging problem in computational fluid dynamics (CFD) ~\cite{Wu2013}. 
Several shock detection methods have been developed for hydrodynamic shocks. Pagendarm and Seitz ~\cite{Pagendarm1992} proposed a shock detection method based on searching the maxima of the density gradient. This method is easy to understand, but proper filters must be designed to remove false results. Samtaney~\cite{samtaney2000} proposed the zero-crossing of the laplacian of the density field to detect shocks. 
Since the normal direction of shock wave is parallel to local pressure gradient, and thus normal Mach number can be obtained from pressure distribution, Lovely and Haimes ~\cite{lovely1999} proposed that the iso-surface of unity normal Mach number represents detected shock wave surface.
The above methods are generally easy to implement, although their effectiveness and accuracy can still be improved.
Kanamori and Suzuki ~\cite{Kanamori2011aiaa, Kanamori2011} have provided a shock detection method by calculating the critical lines of the vector field of the characteristics. This method can be applied in steady and unsteady 2D and 3D flows. A shock-wave-detection technique for continuum and rarefied-gas flows has been proposed basing on the Schlieren imaging ~\cite{Akhlaghi2017}. The scheme is applicable for any existing two-dimensional flow fields obtained by experimental or numerical approaches. Samtaney \etal ~\cite{samtaney2001direct} proposed a method to extract shocklets in turbulence simulations based on finding the minimum of a cost function defined in terms of  local Rankine-Hugoniot jump conditions.
Recently, Fujimoto \etal ~\cite{fujimoto2018, fujimoto2019} offered an alternative shock detection method by integrating Canny Edge Detection, which is one of the image processing methods to detect edges, and Rankine-Hugoniot relations.
These two methods provide good accuracy but its implementation is more complicated.

In magnetohydrodynamics (MHD), a wave is considered physical only if it satisfies both the viscosity admissibility condition and the evolutionary condition~\cite{falle2001}. In a \emph{strongly planar} system, the fast and slow waves (shocks or expansion fans), contact discontinuities, $1 \to 3$ and $2 \to 4$ intermediate shocks, slow and fast compound waves are considered admissible; while fast and slow waves (shocks or expansion fans), contact discontinuities and $180^{\circ}$ rotational discontinuities (RDs) are admissible in the \emph{planar} system.
Note that a flow is \emph{planar} if there is no derivatives in the out-of-plane $(z)$ direction, and \emph{strongly planar} if there is also a reference frame in which there is no vector component in the $z$-direction.
Therefore, detection of shock waves in MHD becomes more challenging due to the variety of wave types ~\cite{kennel1989, falle2001}. Wheatley \etal ~\cite{Wheatley2005JFM} discuss in detail the variety of waves resulting from shock refraction in MHD. 
Most of the shock detection methods developed for hydrodynamics have not yet been adopted or extended for MHD. 
Traditionally, the classification of MHD waves is based on the Rankine-Hugoniot relations, which lead to a set of nonlinear algebraic relations and compatibility relations for a complex wave structure.  
Snow \etal~\cite{snow2021} proposed detection method of MHD shocks based on their upstream and downstream velocity relative to the characteristic speeds of the system. This method has been specifically applied to shock identification and classification in 2D MHD compressible turbulence-Orszag-Tang vortex evolution. 
Another example is that of chromospheric detections of intermediate shocks for MHD nonlinearities in sunspot atmospheres~\cite{Houston2020} and a technique to identify the high-frequency MHD waves of an oscillating coronal loop~\cite{allian2021} has been developed. 
It is claimed that methods for the detection of MHD waves detection are still somewhat limited, and a more thorough investigation of MHD waves detection is still an unexplored area. 
\begin{figure}[ht]   
\centering
\subfigure[]{
\includegraphics[scale=0.13]{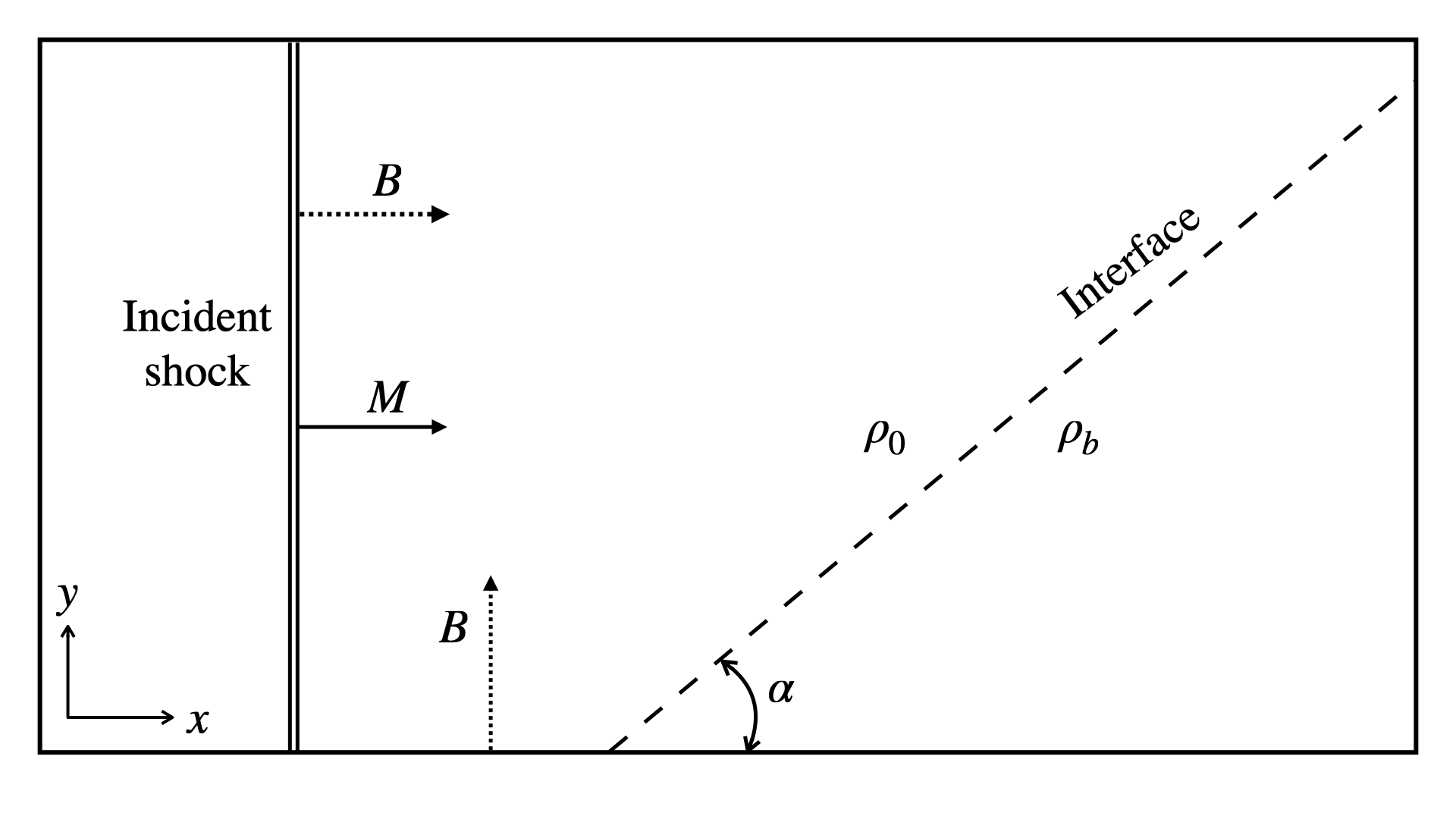}\label{fig01_a}
} 
\subfigure[]{ 
\includegraphics[scale=0.7]{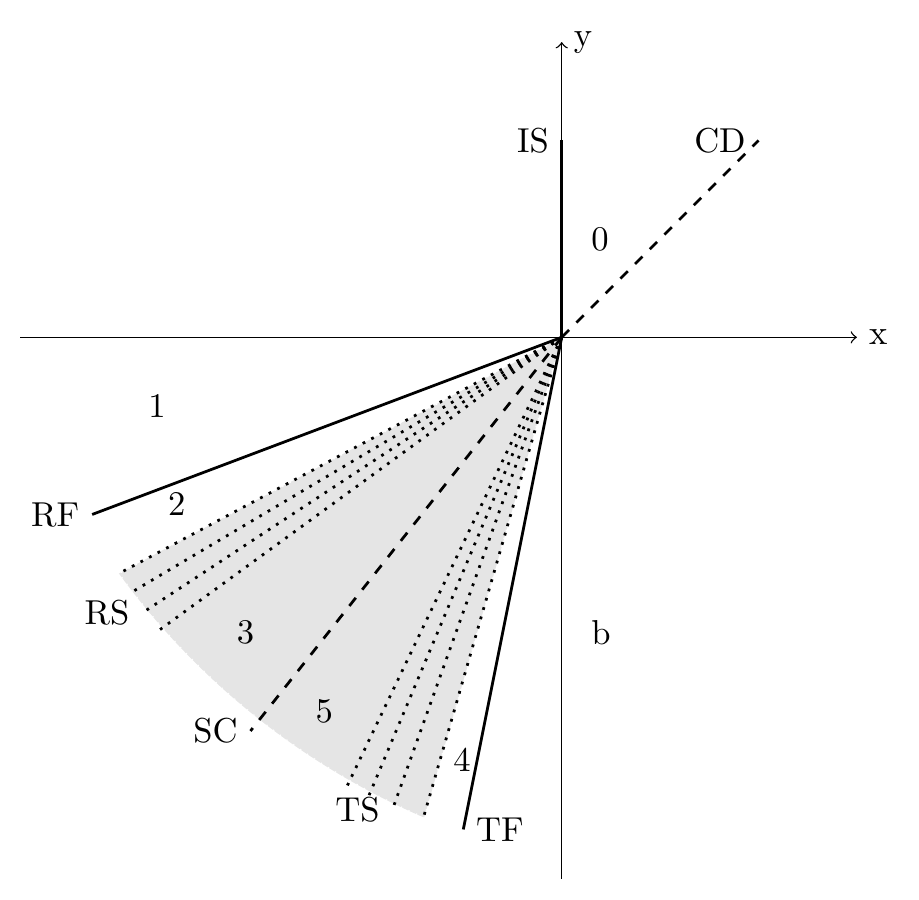}\label{fig01_b}
} 
\caption{(a): Physical set-up for the problem of MHD shock refraction. (b): A wave structure of MHD shock refraction with initially presence of $B_y$ and $(M=2, \eta=3, \gamma =1.4, \alpha = \pi/4, \beta = 2 )$, $IS$, incident shock; $CD$, contact discontinuity; $TF$, transmitted fast shock; $TS$, transmitted slow-mode expansion wave; $SC$, shocked contact; $RS$, reflected slow-mode expansion; $RF$, reflected fast wave.}
\label{fig01}        
\end{figure} 

\begin{figure}[ht] 
\includegraphics[scale=0.8]{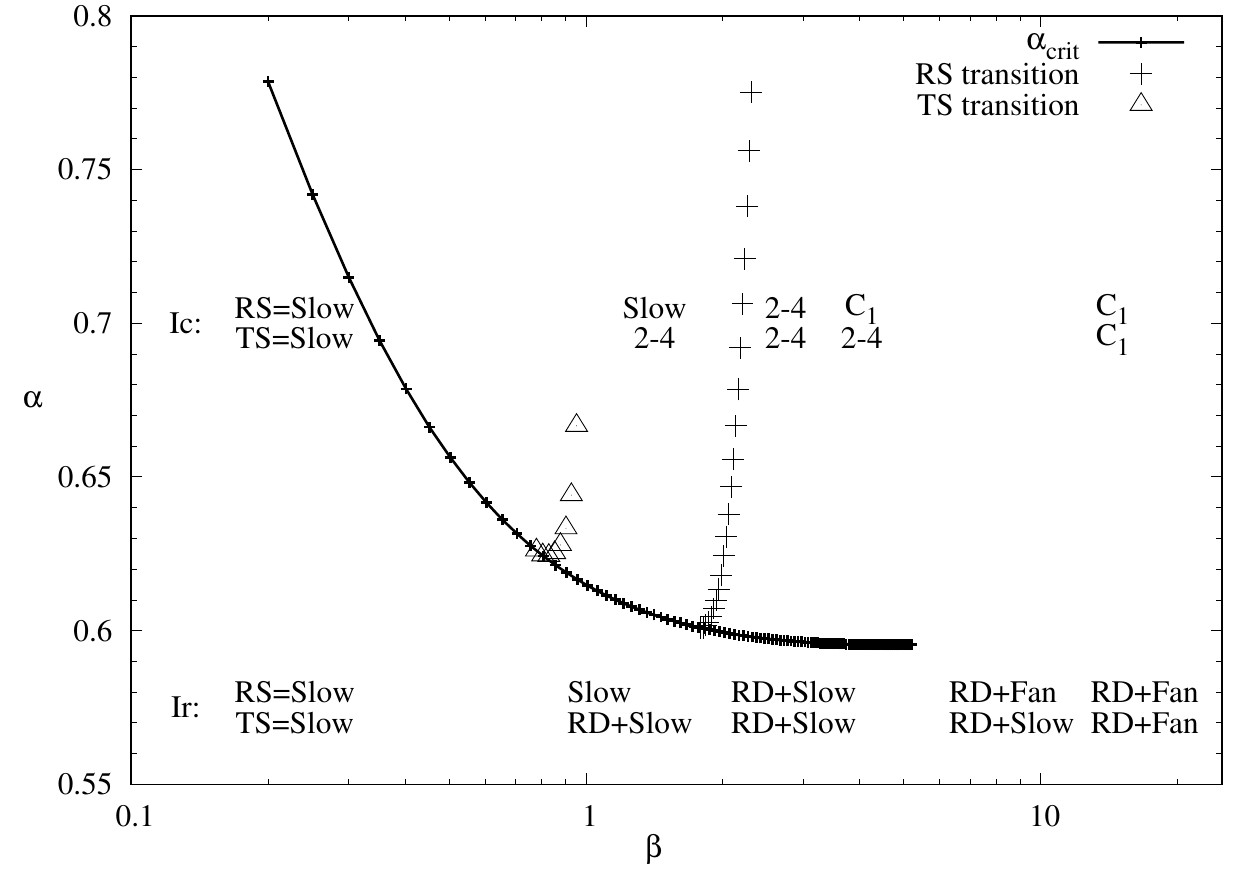}
\caption{\label{fig02} Locations of transitions in solution type with increasing $\beta$ ($B_x$ is initially present) along Branches $Ic$ and $Ir$ $(M=2, \eta=3, \gamma =1.4)$ ~\cite{Wheatley2005JFM}. $2-4$ designates a $2 \to 4$ intermediate shock, $C_1$ designates a slow-mode compound wave, $RD$ and $Fan$ designate a rotational discontinuity and an expansion fan, respectively. Branch $Ic$ $(Ir)$ denotes strongly planar system (planar system).} 
\end{figure}

Presently, we develop a simple method to detect MHD shock waves based on machine learning. Our method is applied to MHD shock refraction at an inclined density interface (see, for example, references ~\cite{Samtaney2003} and ~\cite{Wheatley2005JFM}). 
A canonical physical set-up to investigate shock refraction is shown in Fig.~\ref{fig01_a}. 
The flow is characterized by the following parameters: the incident hydrodynamic shock sonic Mach number $M$ (or the fast magnetosonic Mach number for fast mode MHD shocks), the density ratio of the interface $\eta = \rho_b / \rho_0$, the ratio of specific heats $\gamma$, the angle between the incident shock normal and interface $\alpha$, and the non-dimensional strength of an initially applied magnetic field $\beta^{-1} = B^2/{2p_0}$, where $B$ and $p_0$ denotes the dimensionless magnitude of the applied magnetic field and the pressure of the gas.  
The shock refraction process results in a pair of reflected waves and a pair of transmitted waves as shown in Fig.~\ref{fig01_b}. $RF$ and $TF$ are the fast reflected and transmitted magneto-sonic shocks, respectively;  whereas the wave type of $TS$ and $RS$ is strongly dependent on the chosen parameters (\eg $\beta, \alpha$ and initial magnetic field orientation whether parallel or perpendicular to the motion of incident shock, etc.) ~\cite{Wheatley2005JFM, Chen2022}. Fig.~\ref{fig02} shows an example of $RS$ and $TS$ transitions with increasing $\beta$ ($B_x$ is initially present) by fixing a set of parameters $(M=2, \eta=3, \gamma =1.4)$. 
In the studied range of $\beta$, $RS$ and $TS$ transition from a slow shock to a $2 \to 4$ intermediate shock (resp. $RD$ followed by slow shock) and then into a slow-mode compound wave $C_1$ (resp. $RD$ followed by slow-mode expansion) as $\beta$ is increased in the \emph{strongly planar} system (resp. \emph{planar} system) denoted as $Ic$ ($Ir$). 
It illustrates that, with different parameters, there are many different MHD shock refraction structures which are  too complicated to identify by simply extending the existing methods used for hydrodynamics. 

We approach the shock detection problem by adopting machine learning. 
From the point of view of machine learning application to fluid mechanics, deep learning has achieved some success in CFD in the past few years ~\cite{Colvert2018, Brunton2020, Raissi2020, Novati2021}.
Some shock detection methods based on convolutional neural network (CNN) have been recently proposed ~\cite{Mathew2017, Liu2019} for hydrodynamic cases, and show that the proposed CNN methods produce better detection results than some existing traditional methods.
However, the application of deep learning to MHD is still limited to date. Kates-Harbeck \etal developed a fusion recurrent neural network (FRNN) to predict disruptive instabilities in controlled fusion plasmas ~\cite{Kates-Harbeck2019}. 
It shows a good efficiency of the deep learning application to predict disruptions in fusion reactors. 
The objective of the present work is to develop a simple neural network to detect MHD waves, constraining these waves to the admissible kind for both \emph{planar} and \emph{strongly planar} systems.

The outline of this paper is as follows. In Sec.~\ref{sec:2} we first introduce the fixed output model in which the wave number of each training data is fixed, followed by the detection model without fixing the number of waves for each sample. 
Sec.~\ref{sec:3} presents results and discussions. The two proposed models are first verified via hydrodynamic cases to demonstrate their robustness and efficiency in Sec.~\ref{sec:3a}. 
The application to MHD wave detection in 1D and the reconstruction of 2D shock refraction are discussed in Sec.~\ref{sec:3b} and Sec.~\ref{sec:3c}, respectively. 
Finally, some conclusions are presented in Sec.~\ref{sec:4}.
\section{\label{sec:2} 1D Convolutional Neural Networks}
In this section, we present two CNN algorithms to detect the MHD waves resulting from the shock refraction process. 
The two algorithms are developed based on supervised learning that uses a training dataset to teach models to yield the desired output. 
It implies that we need to provide a labeled dataset to train algorithms to classify data and predict outcomes accurately.
The labeled dataset consists of input training datasets and labels. 
The primitive variables $(\rho, p, \boldsymbol{V}, \boldsymbol{B})$ (each such state vector consists of six variables) are employed as input training dataset, where $\rho$ and $p$ is the density and the gas pressure, respectively; while $\boldsymbol{V}$ and $\boldsymbol{B}$ denotes the velocity vector and the magnetic induction each with two components, respectively.
Each primitive variable is specified in a training sample. Each training sample comprises on $N_t$ points. Presently, we choose $N_t=100$ points in each training sample. 
Furthermore, each training sample is obtained from a 1D numerical simulation or a straight line that contains all five waves $RF, RS, SC, TS$ and $TF$ resulting from the shock refraction process. 
Thus, we have each training data of dimension $(6 \times 100)$. 
The labels of each sample consist of the labeled class $c \in C$ and the bounds $(s_1, s_2)$ for each wave. 
It leads to a $3w$ dimension of labels for a sample that consists of $w$ waves. 
Presently, we consider $C$ classes for seven MHD waves, viz.,  contact discontinuity $(CD)$, fast shock $(FS)$, slow-mode expansion $(Sexp)$, slow-mode compound wave $(Scw)$, slow shock $(SS)$, rotational discontinuity + slow shock $(RD+SS)$ and $2 \to 4$ intermediate shock $(I24)$.
For hydrodynamics cases, the $C$ classes comprise of three waves (\ie contact discontinuity $(CD)$, shock wave $(S)$, and expansion fans $(Exp)$). We first test and verify our algorithms in hydrodynamics (see Sec.~\ref{sec:3a}).

\begin{figure}[ht] 
\includegraphics[scale=0.8]{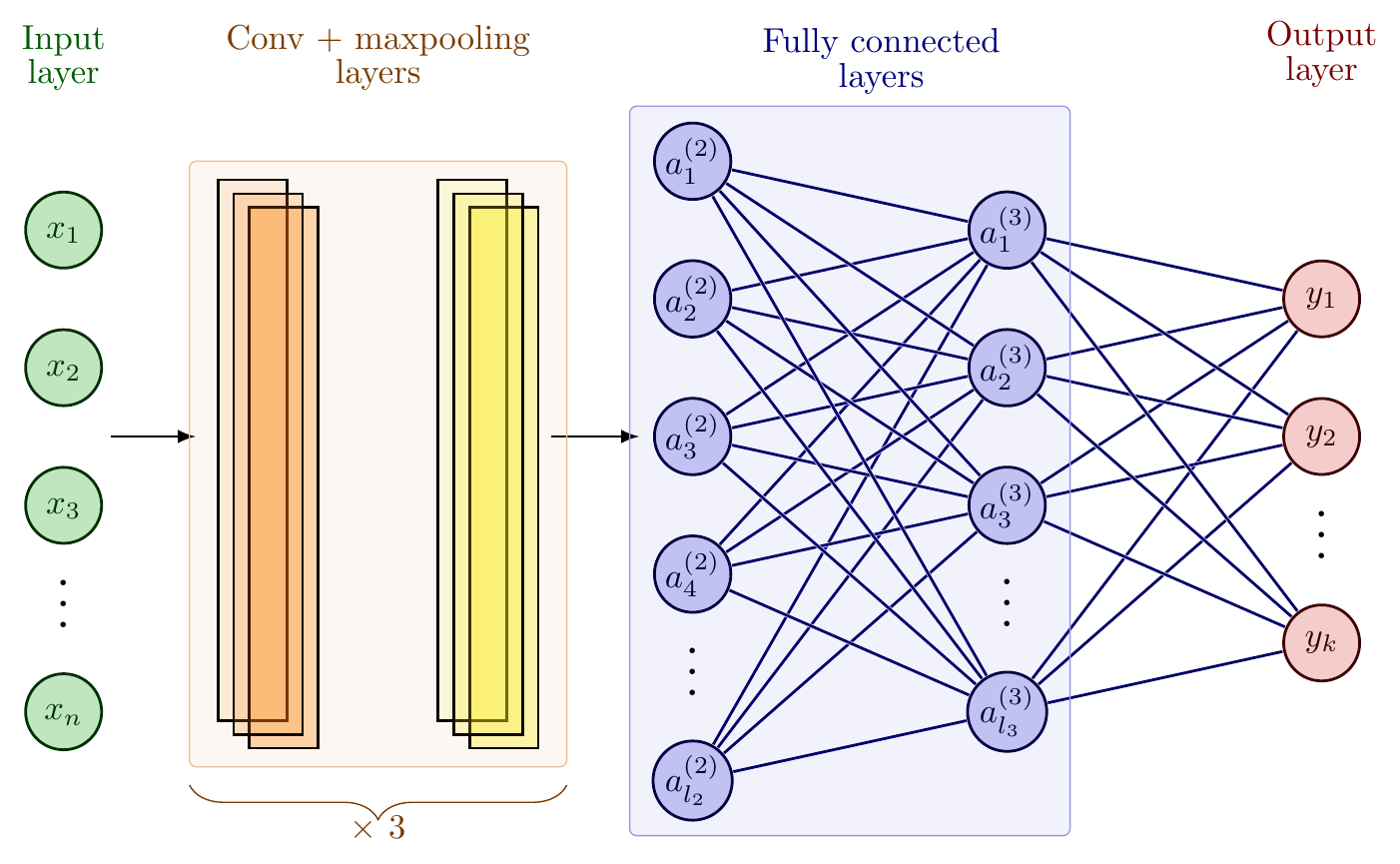}
\caption{\label{fig03} Developed 1D CNN architecture for the MHD wave detection. The network has 3 convolutional and max-pooling layers followed by 2 fully connected layers.} 
\end{figure}

Fig.~\ref{fig03} shows 1D CNN architecture used in the MHD wave detection. 
The developed two CNN algorithms have the same architecture except for the output layer, which is different between the two proposed algorithms. The details of the two CNN algorithms are presented in the subsections Sec.~\ref{sec:2a} and Sec.~\ref{sec:2b}.
The neural network architecture is composed of the input layer, convolutional layer followed by max-pooling layer (3 levels), two fully connected layers, and the output layer. 
The set of some main hyper-parameters of two algorithms are common, \eg the optimizer ``Adam" is chosen since gradient descent algorithms are basic optimizations to attain the optimized value, and the ``Adam" algorithm is one of the most traditional optimizers for its robustness and effectiveness. 
Another hyper-parameter is the activation function ``ReLU" (Rectified Linear Unit) that is set to achieve the nonlinear connection between the fully connected layers. 
The set of the other hyper-parameters of the two algorithms and their details are presented in the following subsections.

%
\subsection{\label{sec:2a} Fixed output model} 
We now present the first CNN model in which the number of waves $w$ of each training sample is fixed ($w=5$ for MHD , and $w=3$ for hydrodynamic shock refraction in 1D).
The fixed output system models both classification and regression problems.
For each wave, the system leads to a $(2+C)$ tensor for predicting the wave location and classifying its type. The output layer is then encoded as a $w \times (2+C)$ tensor. 
Here, we briefly introduce the settings of hyper-parameters in this model: 
\begin{itemize}
    \item Prediction of $w$ waves' locations leading to $2w$ neurons at the output layer, \ie 10 (6) neurons for MHD (hydrodynamic) wave detection. This is a regression problem and the activation function connected to the previous layer is set as the ``ReLU" type. This output of this part is evaluated by minimizing the cost function ``mse" (mean square error). The metric to monitor the training process is set as ``rmse" (root mean square error).  
    \item Classification of $w$ waves' types. This is a  multi-classification problem ($C$ classes in total) resulting in $w*C$ neurons due to the application of the ``softmax" activation function at the final output layer. The number of neurons from the $C$-classification is 35 for MHD cases, while it is 9 for hydrodynamic cases. The traditional cost function for multi-classification problem ``categorical-crossentropy" is adopted and the ``accuracy" is used as a metric. 
\end{itemize}
The remaining specific values of the hyper-parameters setting for this method are in Appendix ~\ref{Appa}. 
Finally, we point out that this algorithm is not a true detection algorithm since the number of waves is fixed {\it a priori} instead of leaving this unspecified. 
However, this model can serve as a verification process owing to the similarity of its architecture as in the real detection model which will be presented in the following subsection. 

\subsection{\label{sec:2b} Detection model}
In recent times, there are several mature different algorithms that are available for real-time object detection in images, such as Fast $R-CNN$~\cite{girshick2015fast}, $Mask-R-CNN$~\cite{he2017mask} and $YOLO$~\cite{redmon2016}, etc. 
We develop the present detection model inspired from the $YOLO$ algorithm. 
There are two main differences between the present detection model and the previous fixed output model. 
First, we make use of divided grids over the input data and  the number of waves present in the different samples are left unspecified instead of fixing their number. 
On the other hand, this algorithm models detection as a regression problem. In the fixed output model this is regarded as two problems: classification and regression.  
The main idea is as follows. 
\begin{itemize}
    \item The algorithm divides the input data into $N_{grid}$ grids. If the center of a wave falls into a grid cell, then this grid cell is responsible for detecting this wave. Each grid cell predicts $Bo$ bounding boxes (which predict the locations of waves) and confidence scores $\kappa$ for those boxes. Presently, we set $Bo=1$ as this is deemed sufficient for our 1D CNN algorithm. 
    
    \item The confidence score $\kappa$ reflects how confident the model is that the box contains a wave and also how accurate it considers the box that it predicts. We define the $\kappa = Pr(wave) \times IOU^{true}_{pred}$, where $Pr(wave)$ is the probability of the presence of a wave, and $IOU^{true}_{pred}$ denotes the intersection over union $(IOU)$ between the predicted box and the ground truth.
    If no wave exists in that grid cell, $\kappa$ should be zero. Otherwise, the $\kappa$ is equal to the value of $IOU$. 
    
    \item Each bounding box consists of three predictions: two predictions are the bounds of the predicted bounding box denoted as $(s_1, s_2)$ and the third prediction is the confidence score $\kappa$. The $\kappa$ represents the $IOU$ between the predicted bounding box and any ground truth box.
    
    \item Each grid cell also predicts $C$ conditional class probabilities, $Pr(Class_i |wave)$. These probabilities are conditioned on the grid cell containing a wave. We predict one set of class probabilities per grid cell. 
    
    \item We multiply the conditional class probabilities and the individual box confidence predictions as 
    \begin{equation}
        \label{eq01}
        Pr(Class_i |wave) \times Pr(wave) \times IOU^{true}_{pred} = Pr(Class_i) \times IOU^{true}_{pred}.  
    \end{equation}
    This score encodes both the probability of that class appearing in the predicted box and how well the box fits the wave. Thus, we consider that a prediction is right if the classification $Pr(Class_i)$ is correct and $IOU > 0.5$, for example.     
    
\end{itemize}

This algorithm leads to the final output layer encoded as a $N_{grid} \times (3~Bo + C)$ tensor. For the chosen default set of parameters, $Bo=1$ and $N_{grid}=10$, the number of neurons in the output layer is 100 (60) for the MHD (hydrodynamic) case.
Since the problem is modeled as a regression algorithm, the activation function for both fully connected and output layers is set as ``ReLU" type, loss function and metric is ``mse" and ``accuracy", respectively.  
The remaining detailed hyper-parameters settings of the detection algorithm are in Appendix~\ref{Appb}. 

\section{\label{sec:3} Results and discussions}
\subsection{\label{sec:3a} Verification via hydrodynamics}
\begin{figure}[ht]   
\centering
\subfigure[]{
\includegraphics[scale=0.65]{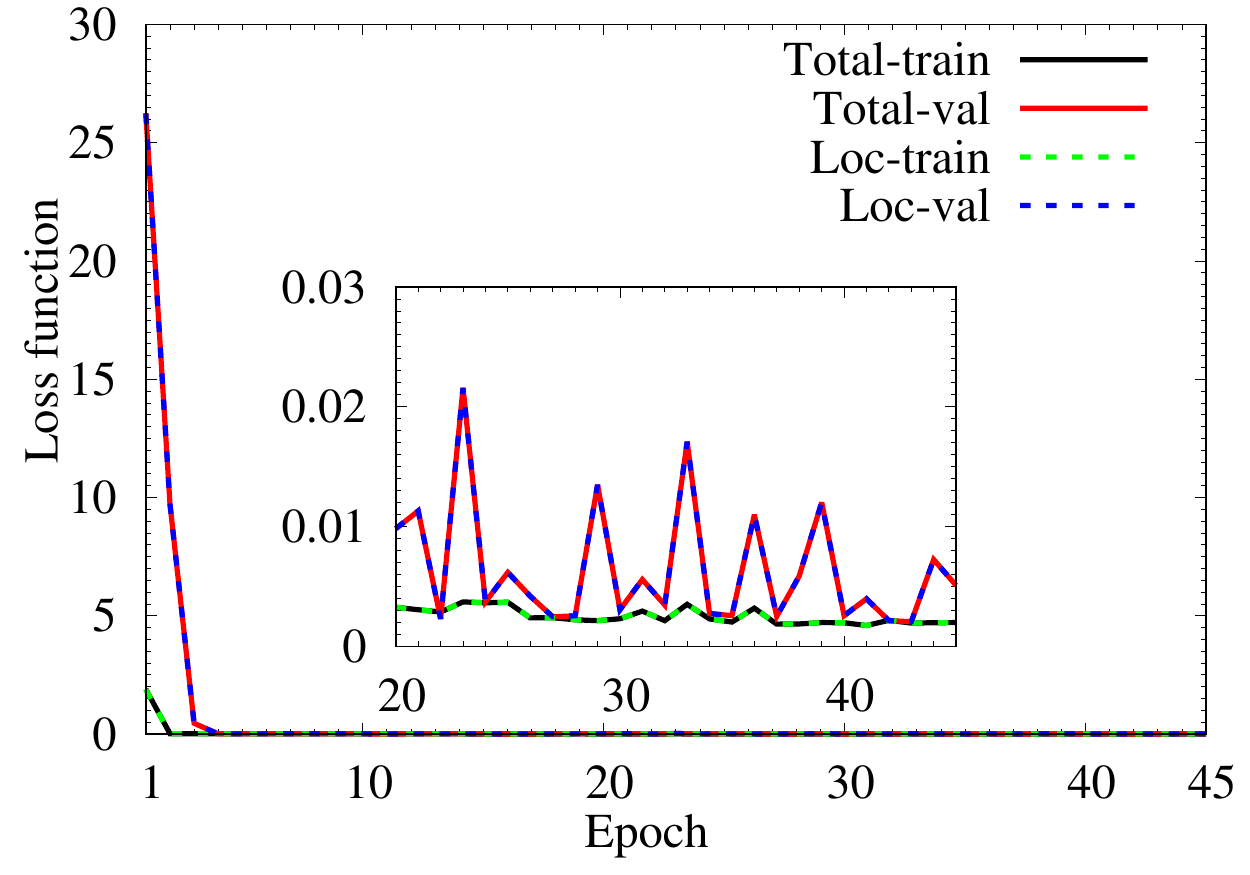}\label{fig04a}
} 
\subfigure[]{ 
\includegraphics[scale=0.65]{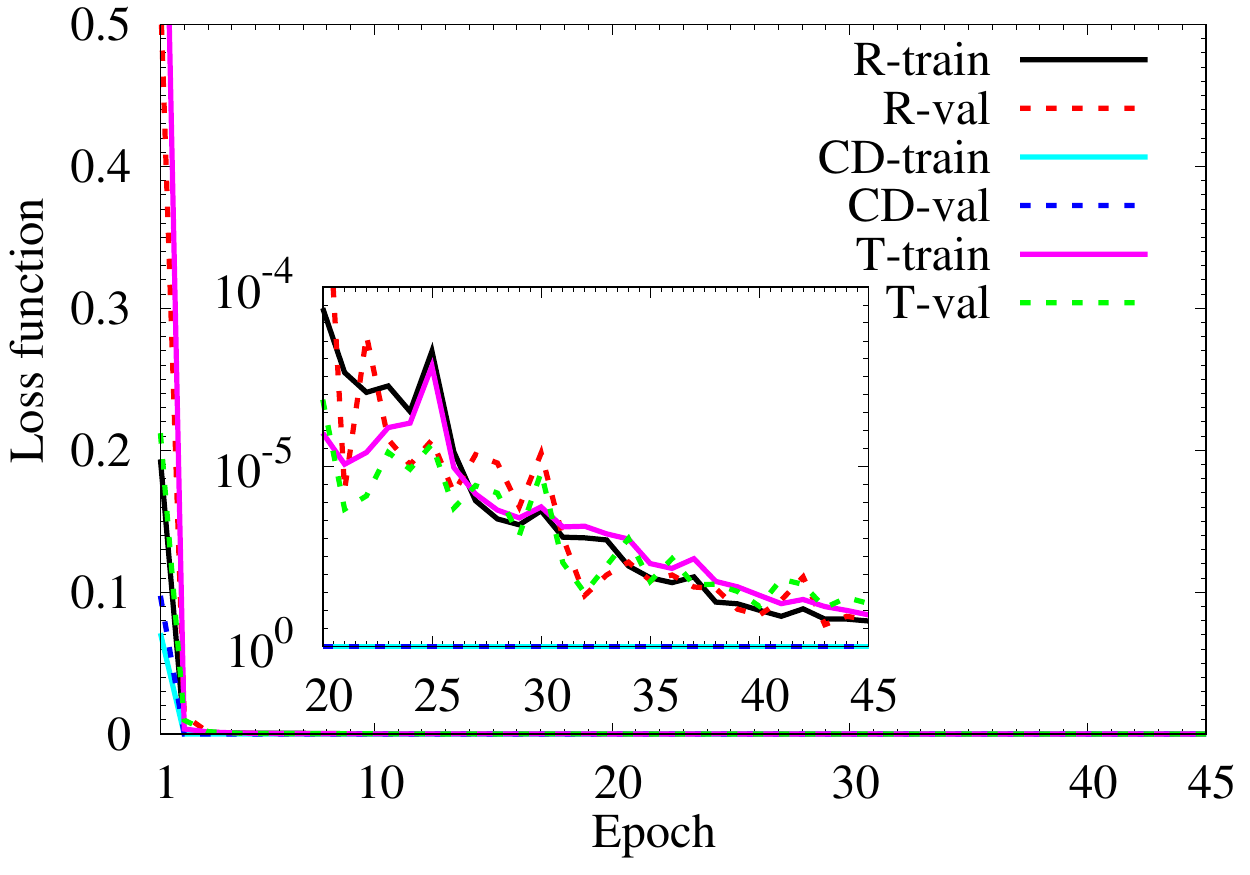}\label{fig04b}
} 
\caption{Metric history of the fixed model, (a): total and location loss function history, (b): loss function of the three separated waves in the 1D hydrodynamic cases.}
\label{fig04}        
\end{figure}

\begin{figure}[ht]   
\centering
\subfigure[]{
\includegraphics[scale=0.65]{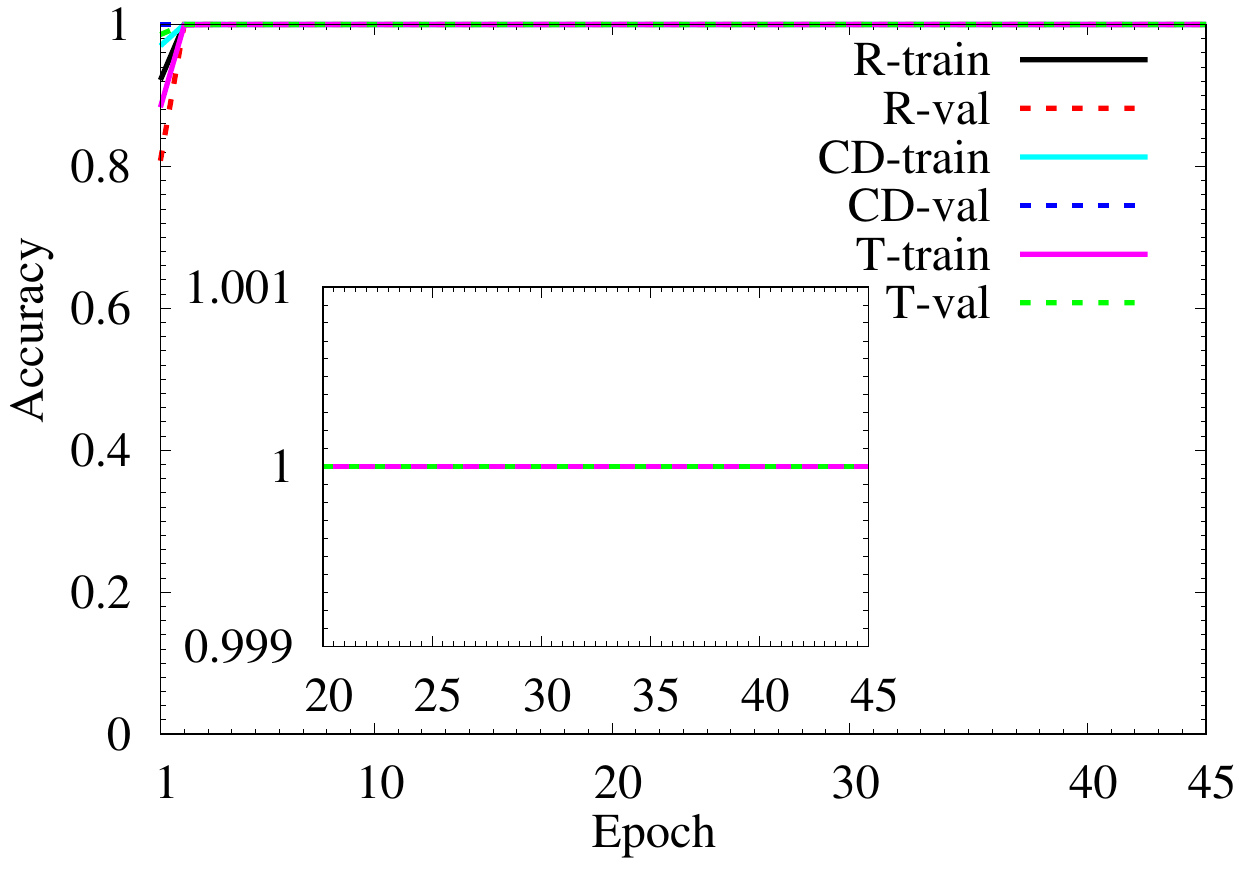}\label{fig05a}
} 
\subfigure[]{ 
\includegraphics[scale=0.65]{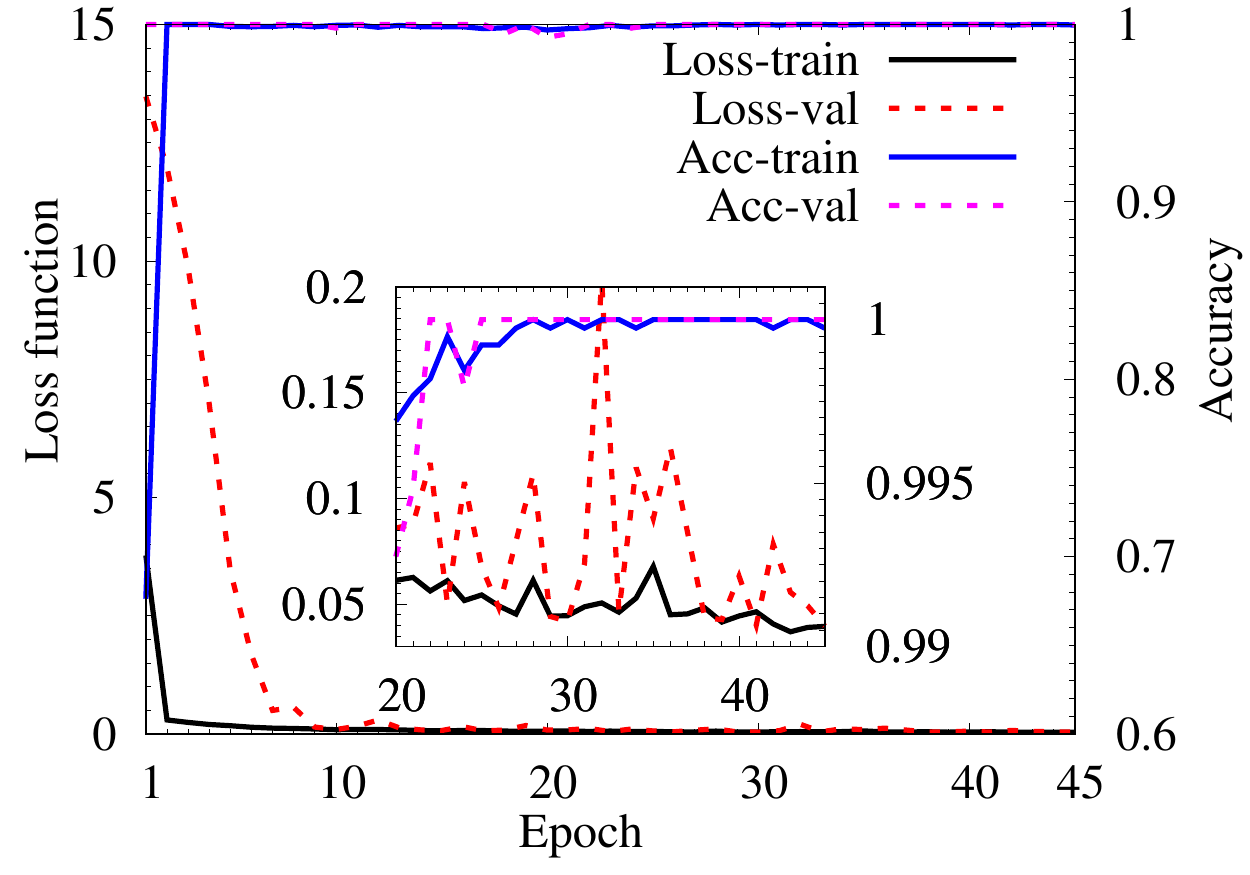}\label{fig05b}
} 
\caption{Accuracy history of (a) the fixed model and (b) the detection model, in 1D hydrodynamic cases }
\label{fig05}        
\end{figure}

We first present verification results obtained from using the above two models for known hydrodynamic cases. 
In hydrodynamics, we choose the training datasets resulting from numerical simulations of 1D shock tube problems~\cite{Sod1978, samtaney1997}. In each sample, the number of waves is fixed: the three waves, scanning the results from left to right in the 1D solution profiles are a reflected shock or expansion fan $R$, a contact discontinuity $CD$ and a transmitted shock $T$.
The detected waves include a combination of shock waves, contact discontinuities and expansion fans. 
Fig.~\ref{fig04} shows the loss function history resulting from the fixed output model.
The loss function history of the wave positions' prediction is almost identical to the total loss function.  
In other words, the main contribution to the loss function comes from the regression part, \ie the prediction of the position of each of the three waves (see Fig.~\ref{fig04a}).
Moreover, the loss function history resulting from the classification is relatively small compared to the prediction one, as shown in Fig.~\ref{fig04b}. 
It leads us to conclude that the fixed output model predicts the waves' types with very high accuracy  (see Fig.~\ref{fig05a}).
The accuracy of classification approaches unity for all three waves. On the other hand, we examine the loss function and accuracy histories from the detection model in Fig.~\ref{fig05b}. 
Since in the detection model we consider the system only as a regression problem instead of the combination of regression and classification problems, the loss function and accuracy histories are plotted as total histories, unlike the separate contributions (prediction and classification) in the fixed output model.
Compared to the fixed output model, the total loss function is relatively more intensive during the phase epoch $>20$, and the accuracy of the detection model is not always exactly unity, but the accuracy is $0.9995$ when the epoch $>30$.
The relatively lower performance of the detection model is attributed to the algorithm unit, which decides if a prediction is correct or not. 
In the detection model, the conditions for deciding whether a prediction is correct or not is monitored by the confidence score within each box and the class probability at each grid cell.  
If the confidence score threshold is set too high, then some waves are not detected, and it results in the decrease of the recall. 
Conversely, some waves may be incorrectly detected if the class probability is set too low, and this leads to a reduction in the precision.    
Therefore, these two thresholds are chosen based on the trade-off between precision and recall.  
Presently, we consider each prediction to be correct only if the confidence score is greater than half ($> 0.5$) and the probability of class $i$ $Pr(Class_i|wave) > 0.5$.
\begin{figure}[ht]   
\centering
\subfigure[]{
\includegraphics[scale=0.6]{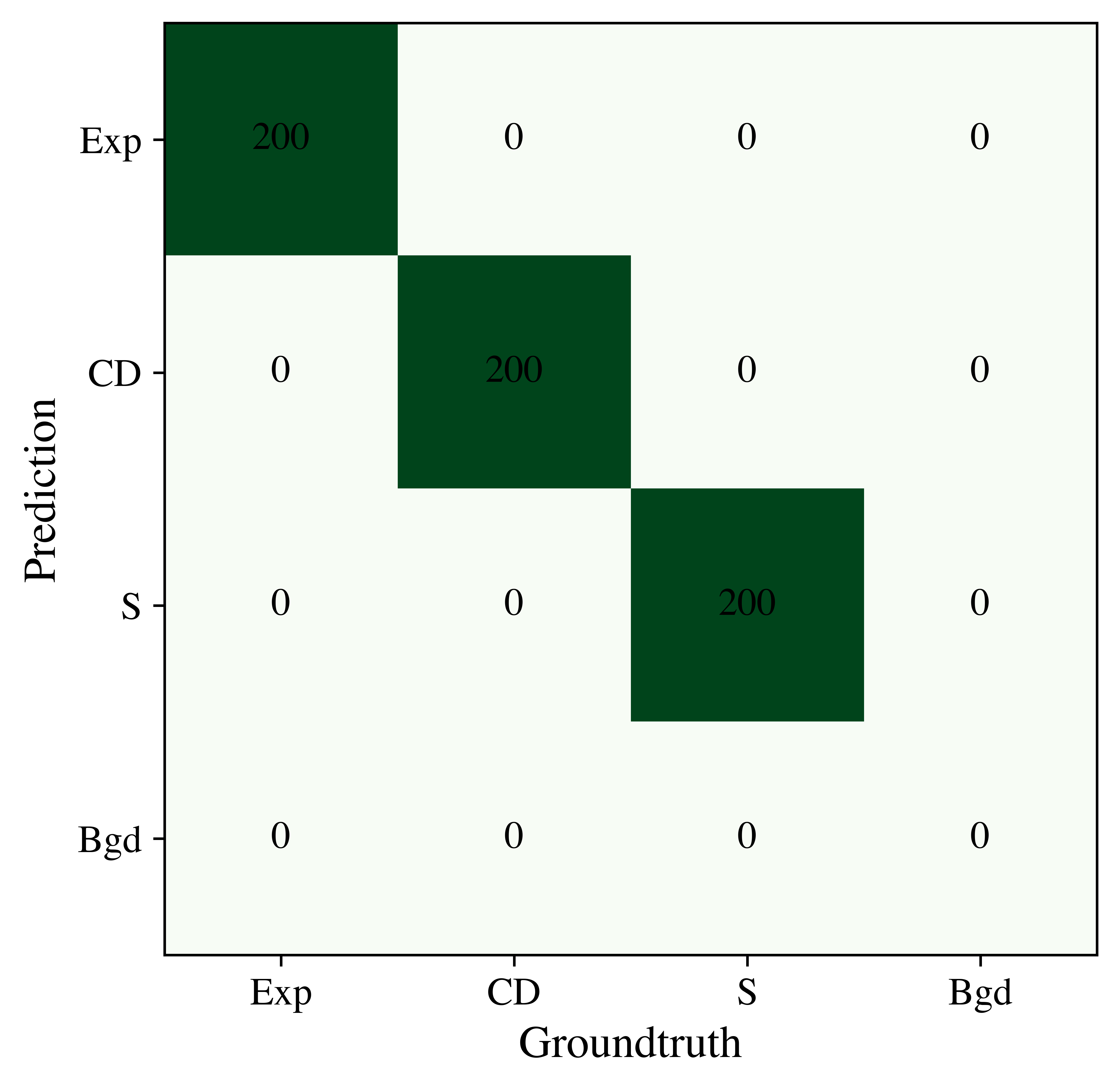}\label{fig06a}
} 
\subfigure[]{ 
\includegraphics[scale=0.6]{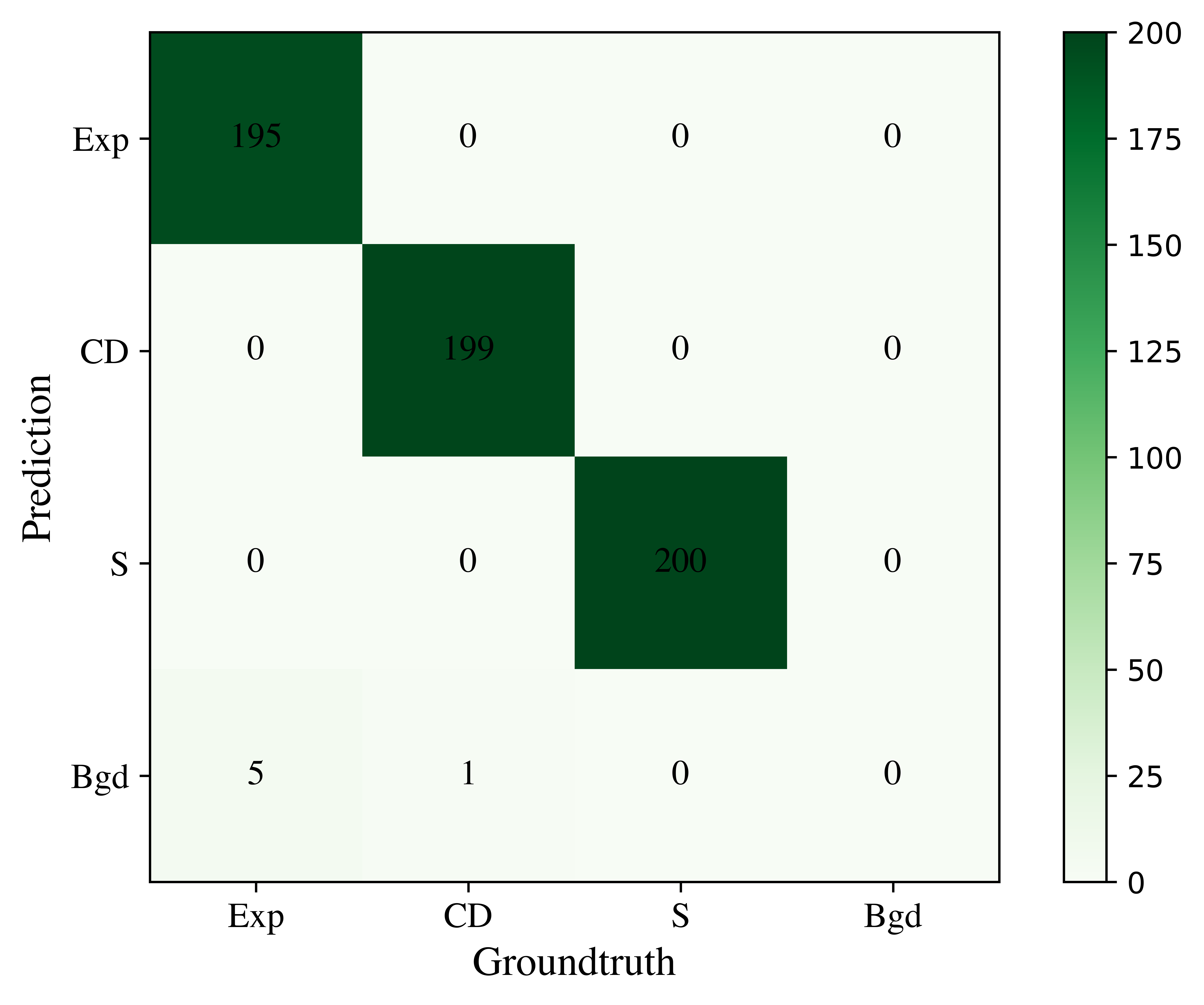}\label{fig06b}
} 
\caption{Confusion matrix of hydrodynamic cases. (a): fixed output model, (b): detection model. $S$, shock; $CD$, contact discontinuity; $Exp$, expansion fans; $Bgd$, background.}
\label{fig06}        
\end{figure} 
After training the models, we employ the two trained models to predict an untrained dataset. The confusion matrices resulting from the two models are shown in Fig.~\ref{fig06}.
Here, we note that the main metrics to analyze predicted results are the accuracy, precision, recall and $F_1$ score, defined as 
\begin{subequations} 
\label{eq02}
 \begin{align}
  & accuracy = \frac{True \ positive + True \ negative}{True \ positive + True \ negative + False \  positive + False \  negative},\label{eq02a} \\ 
  & precision =\frac{True\  positive}{True \ positive + False\  positive},\label{eq02b}\\
  & recall =\frac{True \ positive}{True \ positive + False\  negative},\label{eq02c}\\
  & F_1 =2 \times \frac{precision \times recall}{precision + recall},\label{eq02d}
 \end{align}
\end{subequations}
respectively.   
All the metrics are exactly one for the fixed model, while mean accuracy is $0.99$, mean precision and recall is one and $0.99$, respectively, and $F_1$ is 0.995 for the detection model.
Although the metrics of the detection model are only slightly lower than the ones of the fixed output model, the performance of the detection model is still good.
Thus, we conclude that the detection model may be a promising method to apply in the MHD wave detection.  

\subsection{\label{sec:3b} Detection of MHD waves}
\begin{figure}[ht]   
\centering
\subfigure[]{
\includegraphics[scale=0.65]{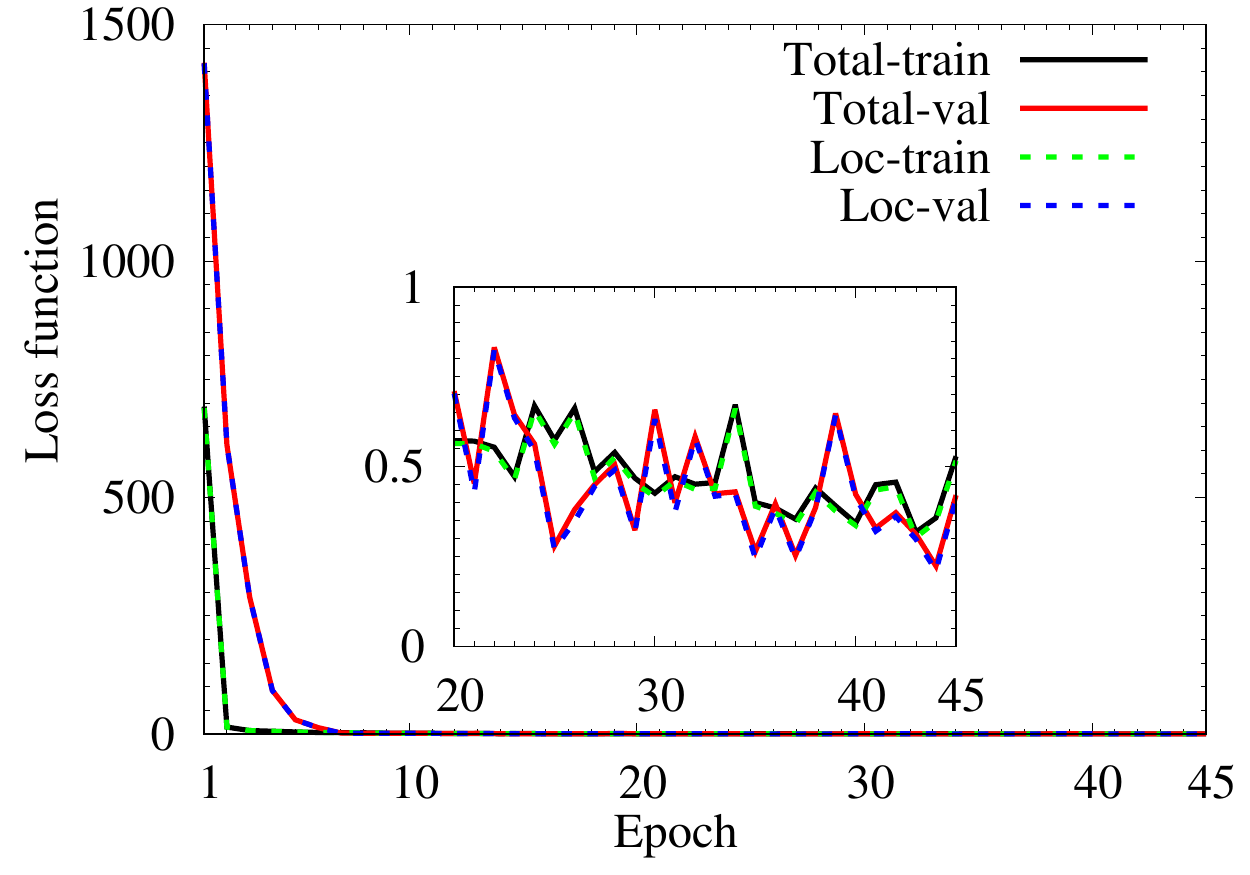}\label{fig07a}
} 
\subfigure[]{ 
\includegraphics[scale=0.65]{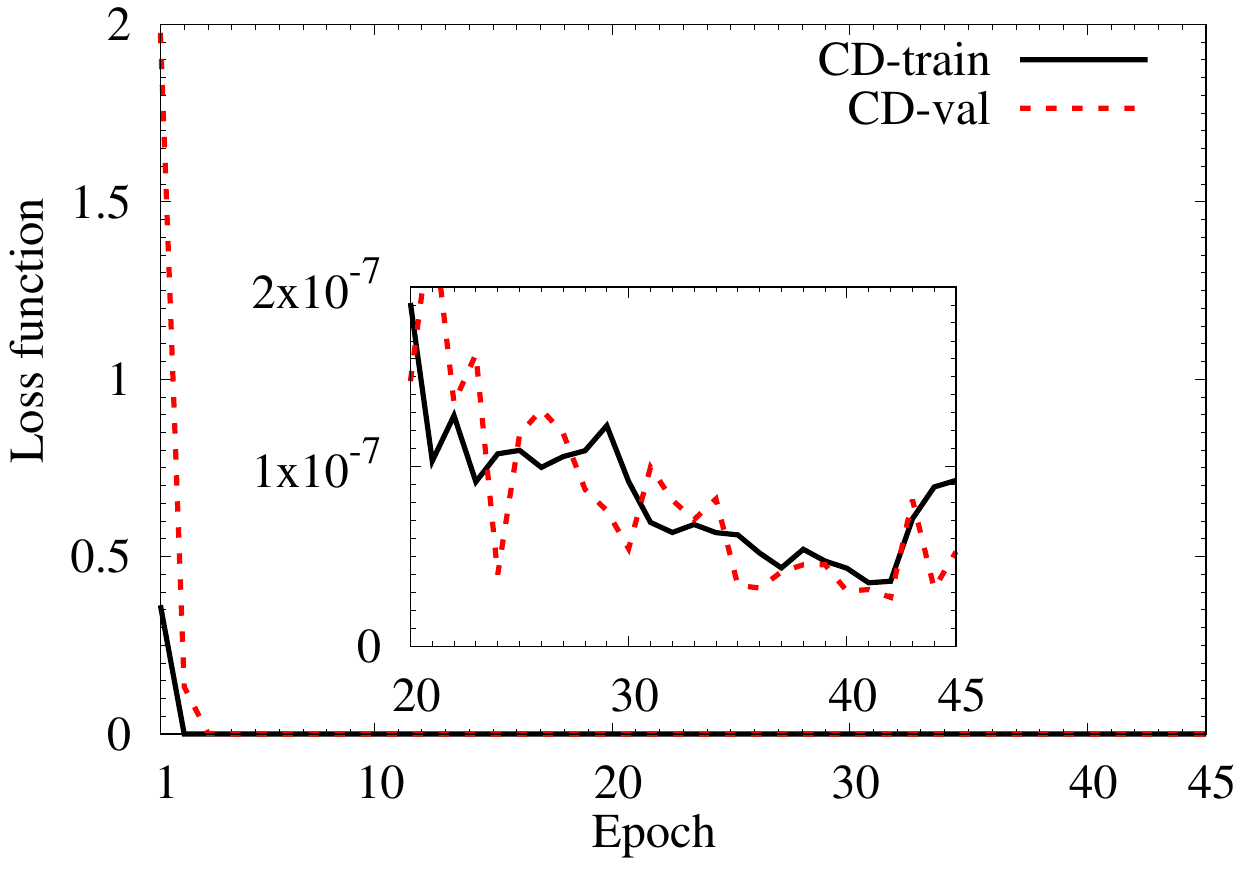}\label{fig07b}
} 
\subfigure[]{ 
\includegraphics[scale=0.65]{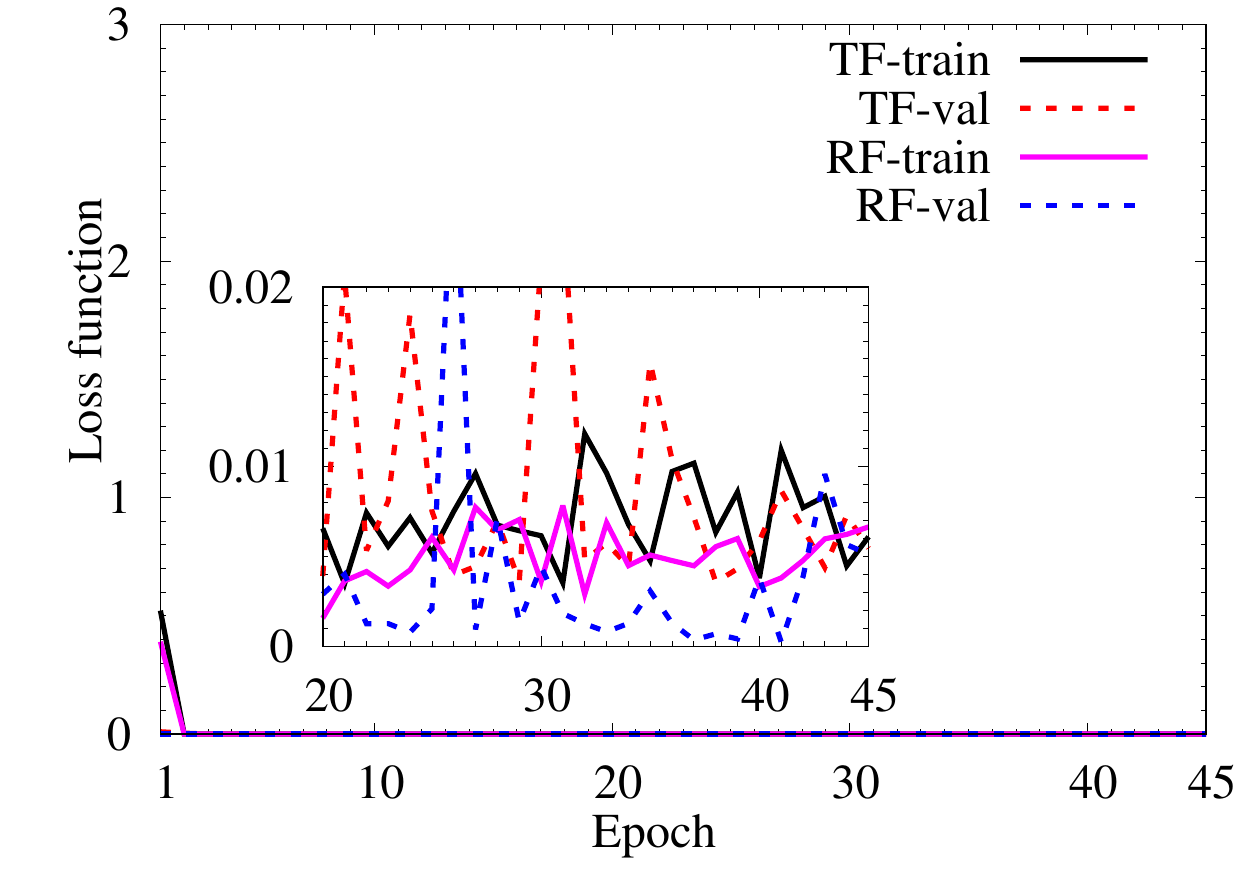}\label{fig07c}
}
\subfigure[]{
\includegraphics[scale=0.65]{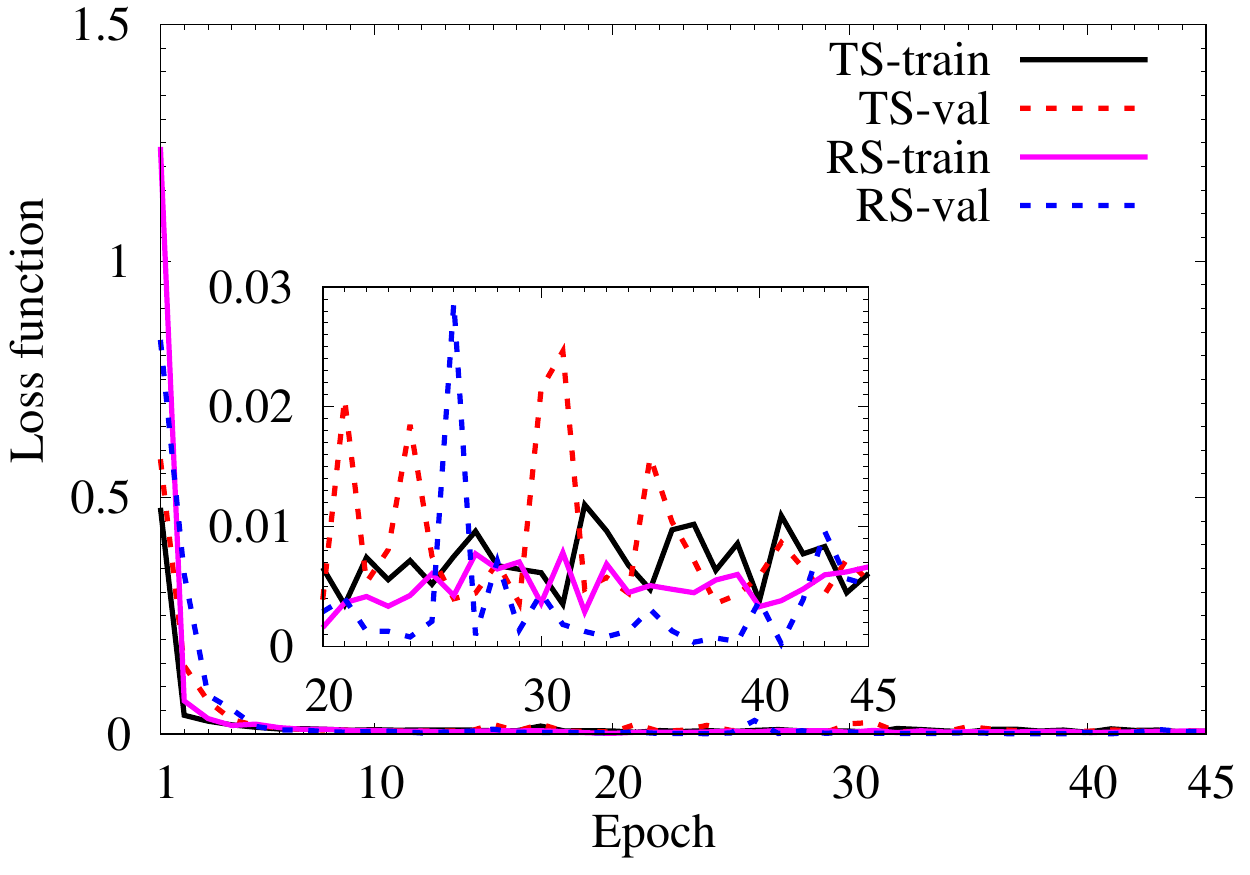}\label{fig07d}
} 
\caption{Loss function history of the fixed output model for MHD cases.}
\label{fig07}        
\end{figure}

\begin{figure}[ht]   
\centering
\subfigure[]{ 
\includegraphics[scale=0.65]{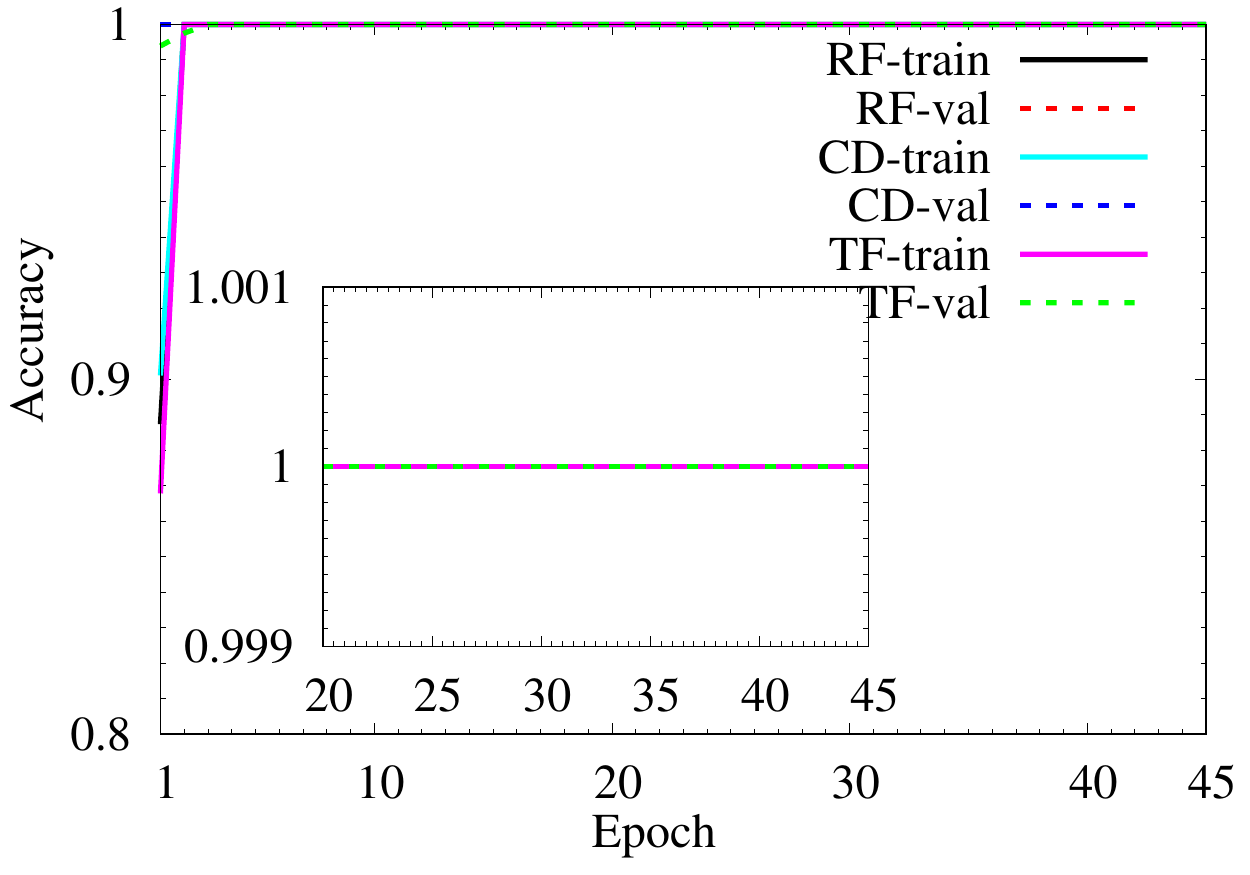}\label{fig08a}
} 
\subfigure[]{ 
\includegraphics[scale=0.65]{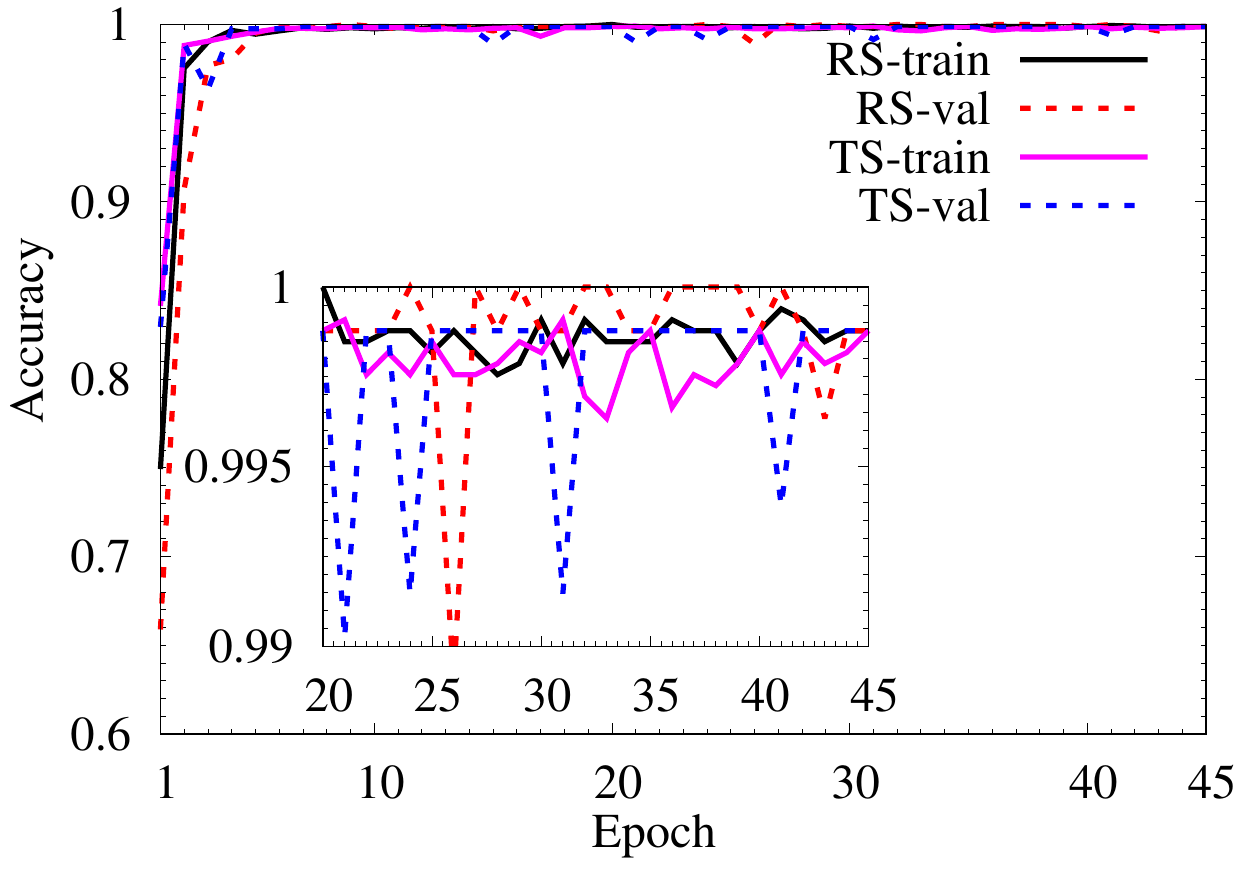}\label{fig08b}
}
\caption{Metric history of the fixed output CNN model for MHD cases.}
\label{fig08}        
\end{figure}

\begin{figure}[ht] 
\includegraphics[scale=0.75]{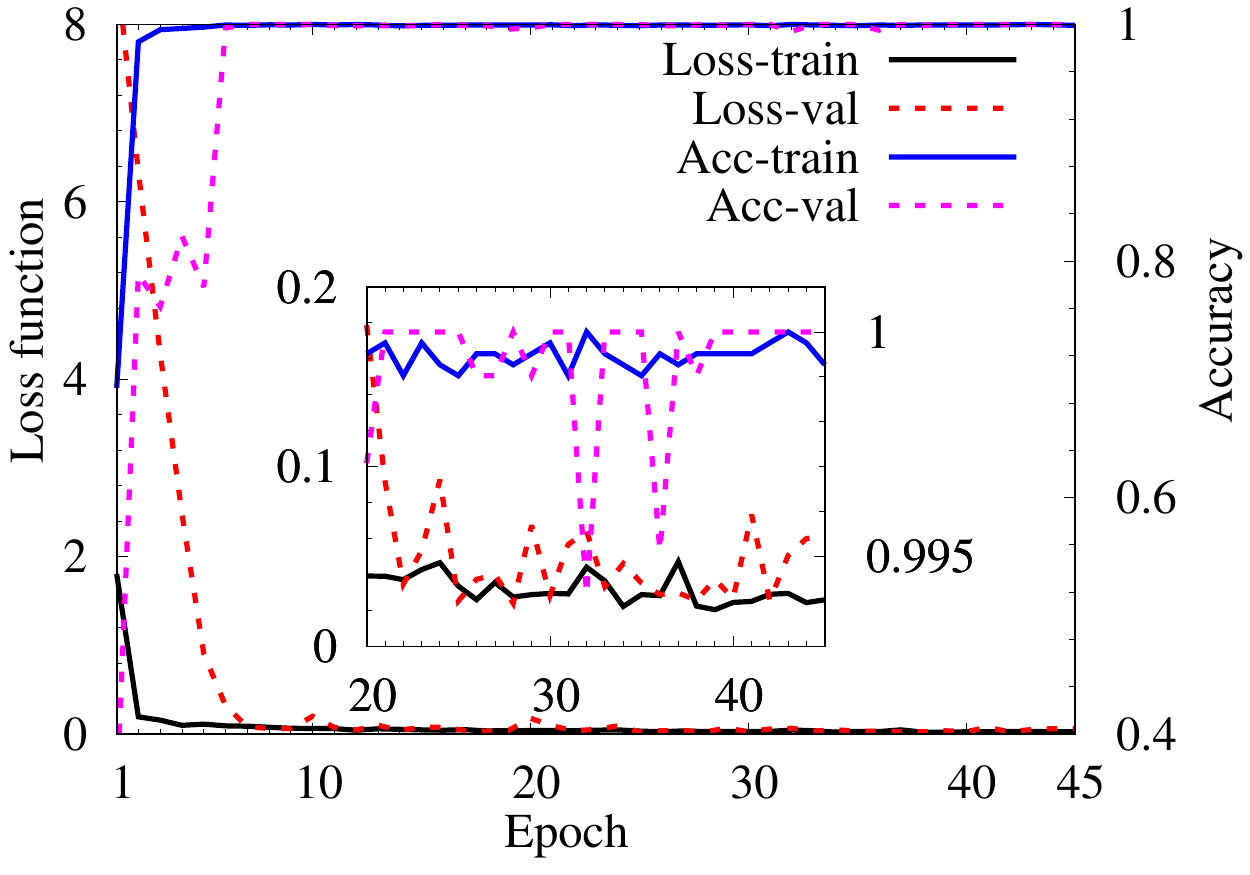}
\caption{\label{fig09} Loss function and accuracy history of the detection model for MHD cases.} 
\end{figure}

\begin{figure}[ht]   
\centering
\subfigure[]{
\includegraphics[scale=0.65]{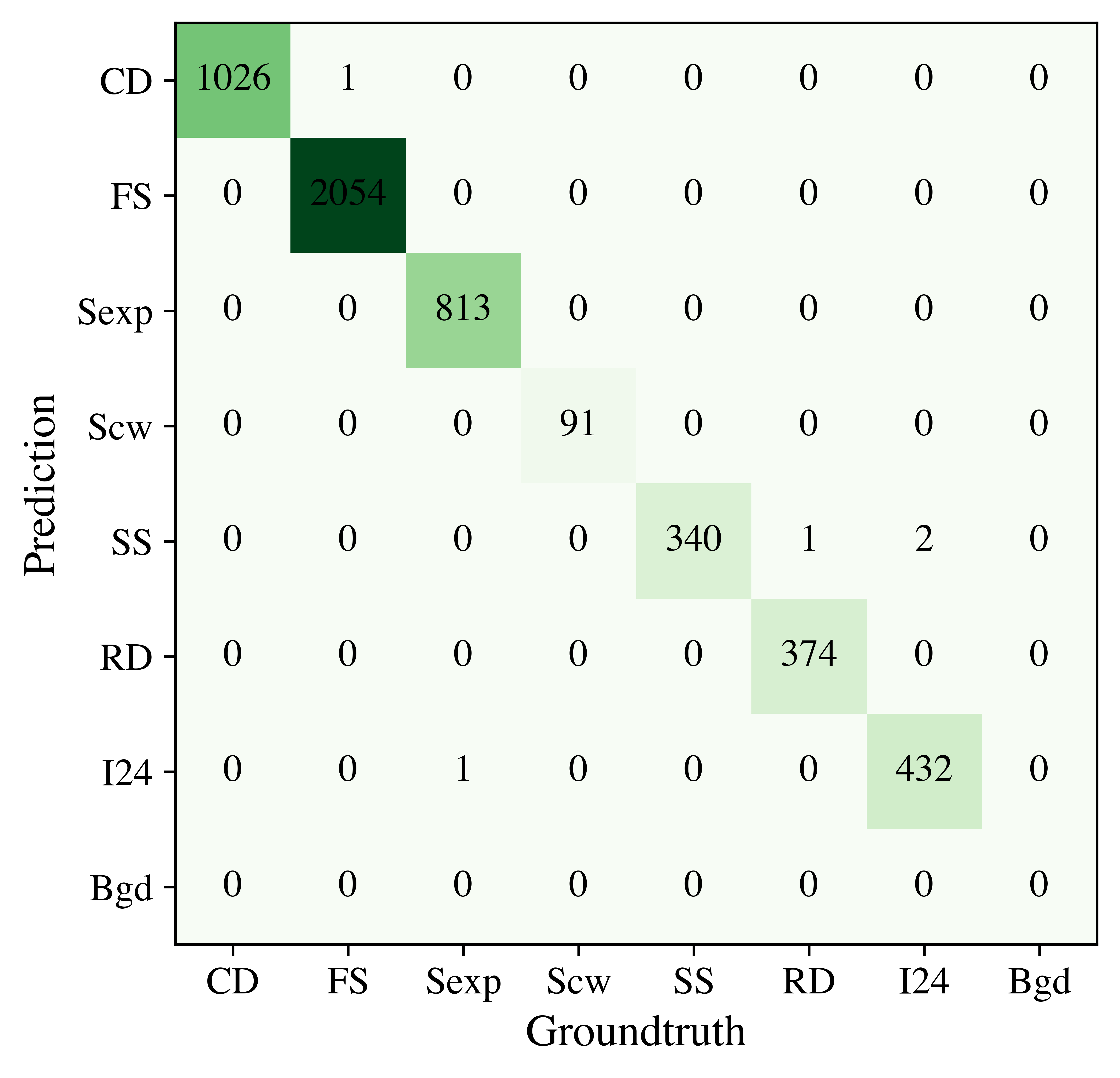}\label{fig10a}
} 
\subfigure[]{ 
\includegraphics[scale=0.65]{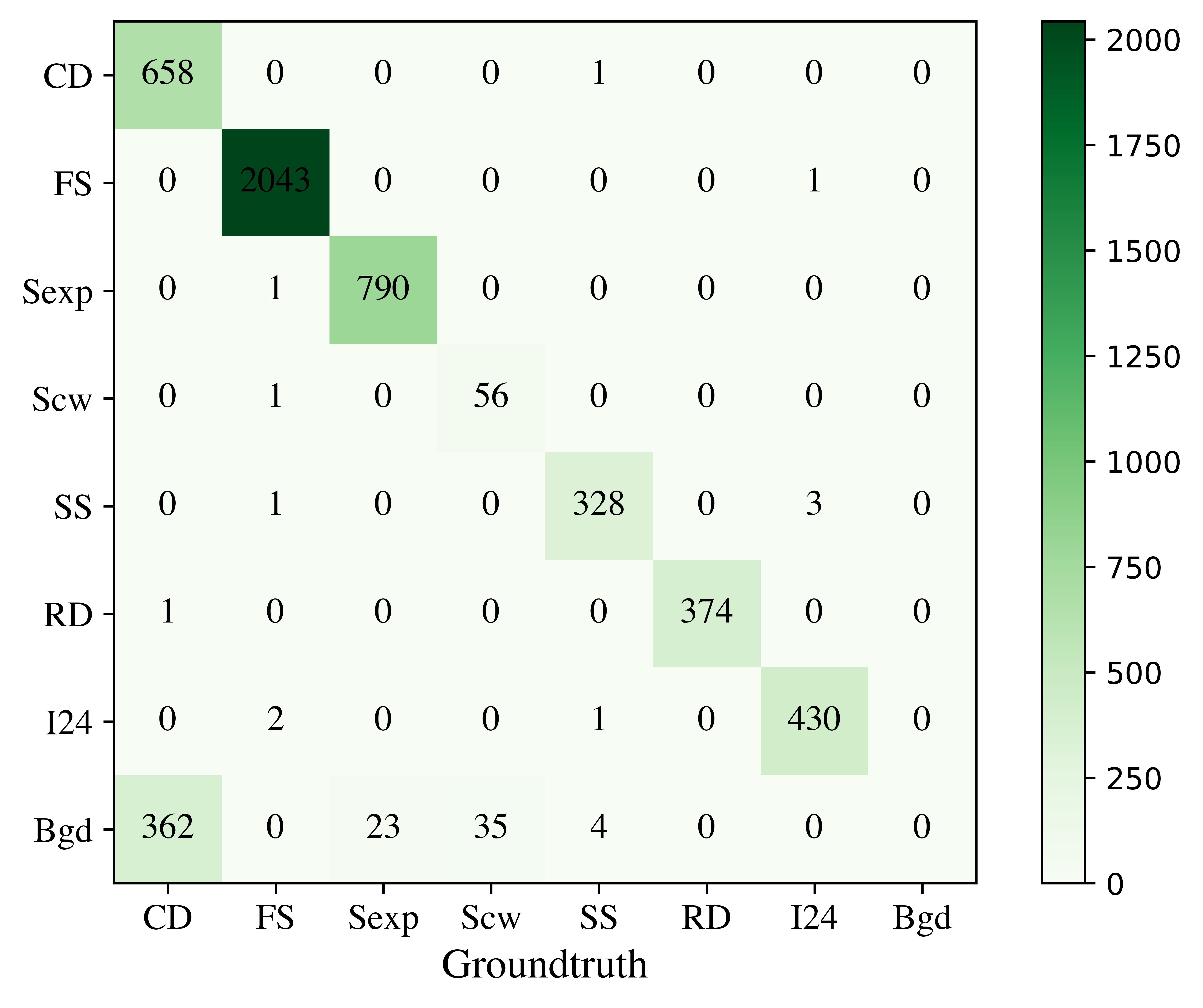}\label{fig10b}
} 
\caption{Confusion matrix of MHD case. (a): fixed output model, (b): detection model.}
\label{fig10}        
\end{figure}

We now examine the detection of MHD waves using the two models that are verified via hydrodynamic cases. 
Here, we use the data resulting from the MHD shock refraction process in 1D as training data.  
There are seven types of MHD waves and five waves $(RF, RS, CD, TS$ and $TF)$ are considered fixed for each sample in the fixed output model, while the number of waves in each sample is not fixed for the detection model. 
As observed in hydrodynamics, the prediction of waves' locations is the main contribution to the total loss function (see Fig.~\ref{fig07a}).
The evolution of total loss function with an increase in the epoch is almost the same as the one of regression (prediction of waves' locations) component. 
For the classification of the middle wave  denoted as CD, the loss function is approximately $10^{-7}$ when the model is close to being well trained (epoch $>20$) (see Fig.~\ref{fig07b}). 
This value is relatively small compared with the others because the middle wave $(CD)$ is shocked contact in all the training datasets.
It is then reasonable that the loss function of $CD$ is the smallest component. 
However, the loss function of classification for the $RF$ and $TF$ waves is about $10^{-2}$ when the model is close to being well trained (see Fig.~\ref{fig07c}).
Unlike the middle wave, each of $RF$ and $TF$ is not of a fixed type: it could either be a fast shock $(FS)$ or slow-mode expansion fans $(S_{exp})$ and this results in a slightly higher loss function at the end of training compared to loss function for the $CD$ wave. 
The same reasoning applies to the evolution of loss function of the classification of the $TS$ and $RS$ waves as shown in Fig. ~\ref{fig07d}. 
$TS$ and $RS$ could either be a slow shock $(SS)$, a $2 \to 4 $ intermediate shock $(I24)$, slow-mode compound wave $(S_{cw})$ or a rotational discontinuity $(RD)$ depending on the difference in the initial conditions. 
Due to the prevalence of multiple wave types in $RS$ and $TS$, the accuracies of $RS$ and $TS$ are not exactly unity at the end of the training phase, while the accuracy is exactly unity for the other three waves $(RF, TF$ and $CD)$ (see Fig.~\ref{fig08}).
Moreover, Fig.~\ref{fig09} shows the loss function and accuracy history resulting from the detection model for MHD cases. 
The loss function tends to be stable and the accuracy approaches $0.999$ at the end of the training phase.
We conclude that the detection model exhibits good performance for the MHD cases. 
The confusion matrices of predicting an untrained dataset resulting from the fixed output model and the detection model are shown in Fig.~\ref{fig10a} and Fig.~\ref{fig10b}, respectively.
For the fixed model, the accuracy is high at $0.999$, and the recall and precision also attain high values:  $0.998$ and $0.999$, respectively. 
This results in a high $F_1=0.998$, which is indicative of the robustness of the fixed model for the detection of MHD waves resulting from shock refraction in 1D. 
On the other hand, a number of waves are not detected by the detection model, especially of the $CD$ type. Recall that, in contrast, the fixed model detects the $CD$ type wave with an accuracy of one and with the smallest loss function. 
Note that the set of parameters is the same as the one for hydrodynamics, $N_{grid} = 10$, the confidence is $0.5$ and the class probability is $0.5$. 
For the detection model, the accuracy is $0.9113$, the recall and precision are $0.9133$ and $0.9975$, respectively, and the $F_1$ score is thus $0.9469$. 
The relative low recall may be attributed to the set of the number of grid cells $N_{grid}$ and the thresholds of the confidence and class probability. 

\begin{figure}[ht]   
\centering
\subfigure[]{
\includegraphics[scale=0.65]{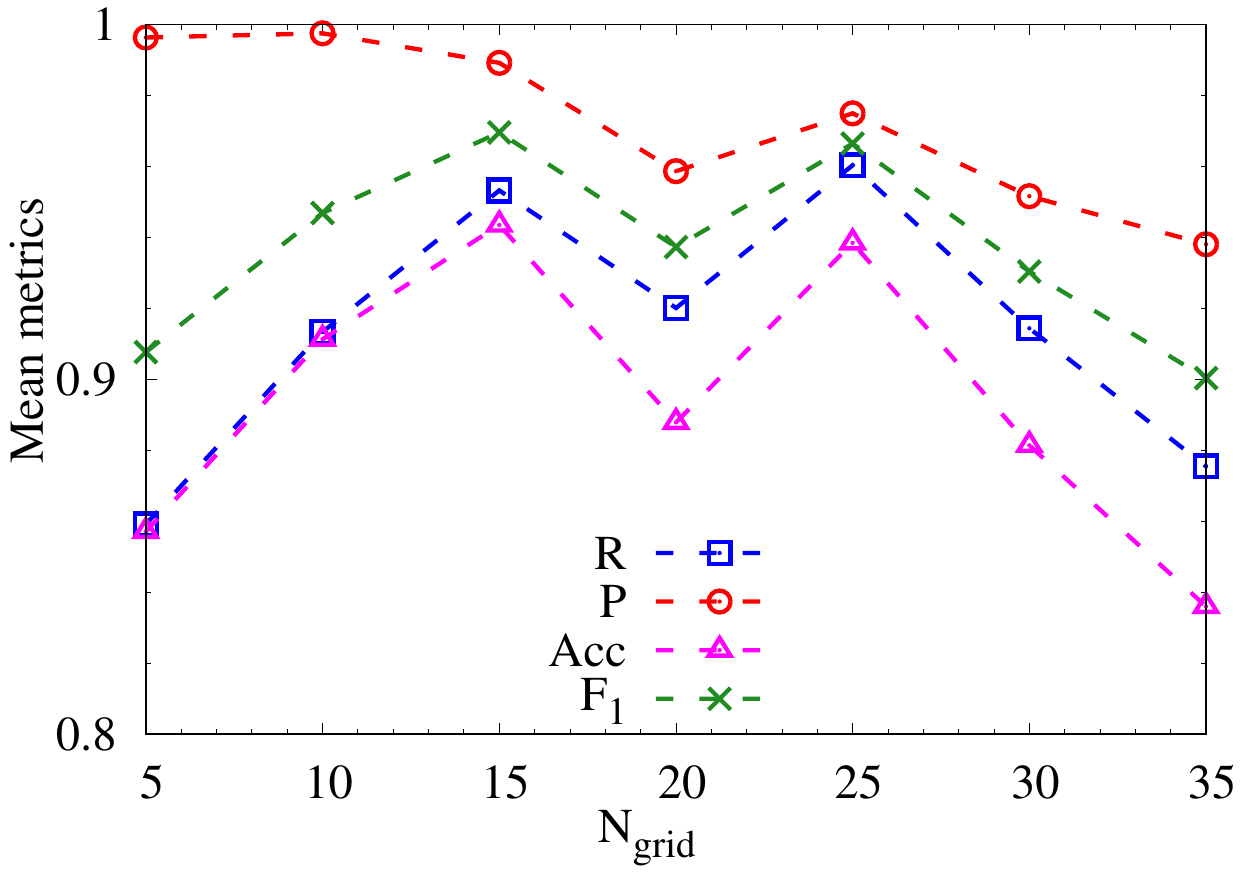}\label{fig11a}
} 
\subfigure[]{ 
\includegraphics[scale=0.65]{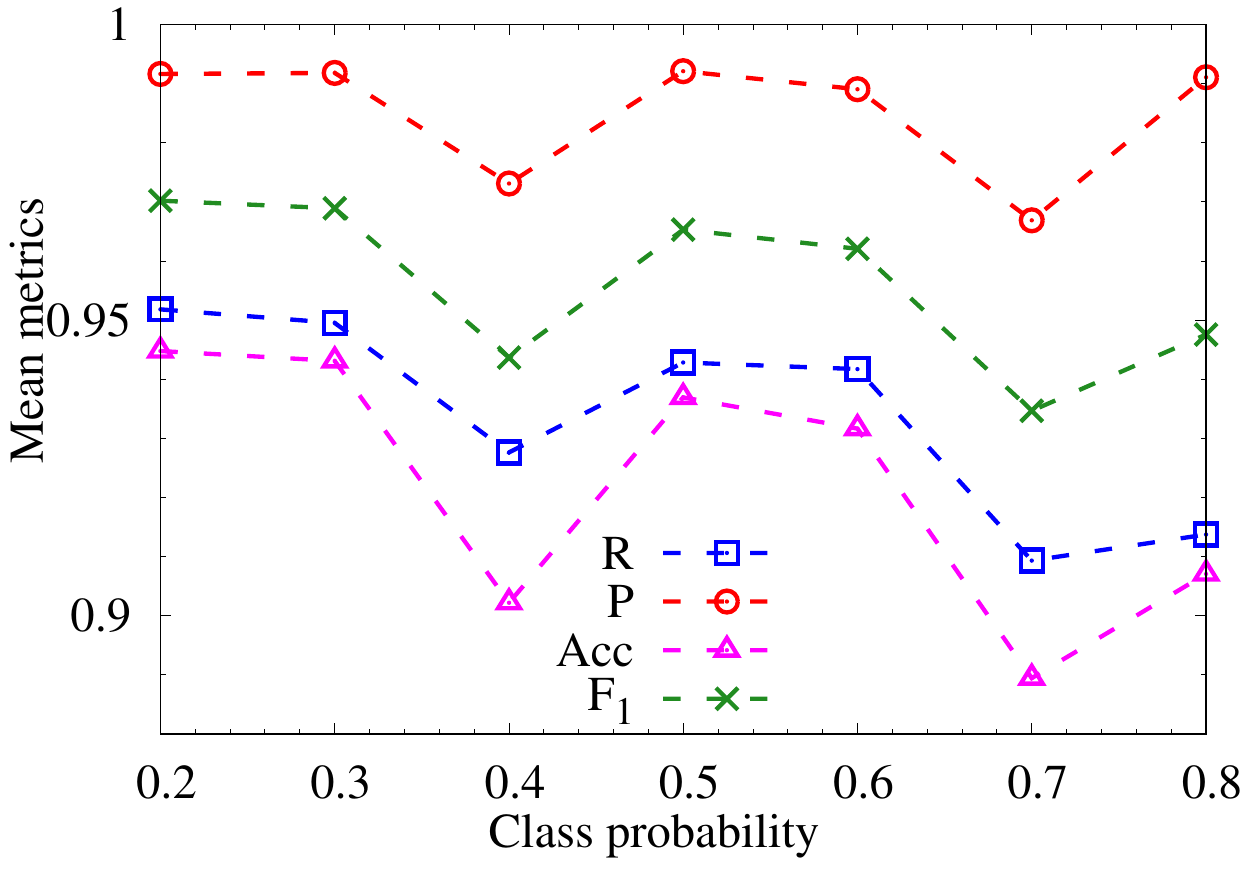}\label{fig11b}
} 
\subfigure[]{ 
\includegraphics[scale=0.65]{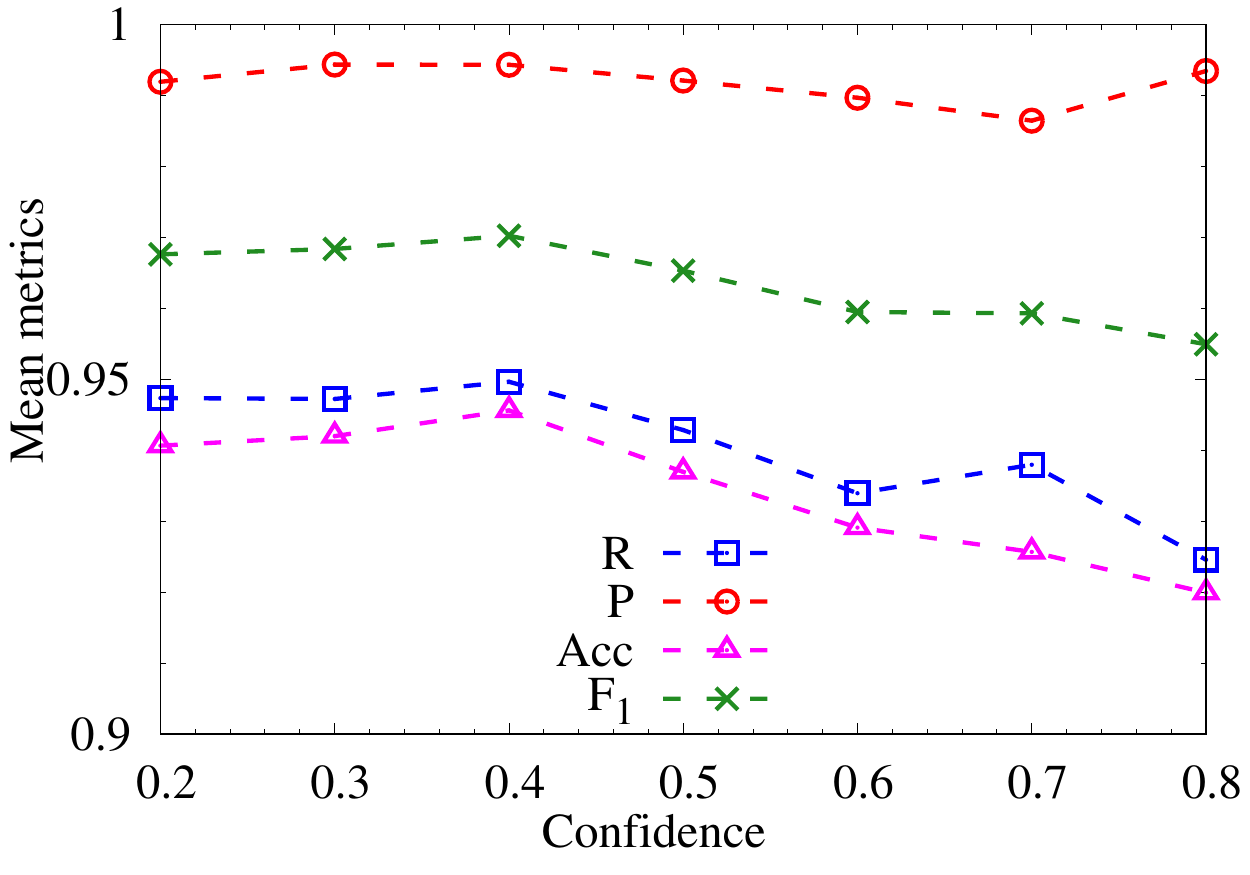}\label{fig11c}
} 
\caption{Mean metrics of the detection model for MHD cases with the different $N_{grid}$ (a), threshold of class probability (b), and confidence (c).}
\label{fig11}        
\end{figure}

The performance of the detection model with different values for $N_{grid}$ is shown in Fig.~\ref{fig11a}.
The evolution of average metrics is not linear as $N_{grid}$ is increased.
In the studied range of $N_{grid}$, all mean metrics have a local minimum at $N_{grid} = 20$, the mean recall, precision and $F_1$ exhibit symmetry about  $N_{grid} = 20$.
It seems that the detection model has a good performance by considering the trade-off between the accuracy ($0.9435$) and $F_1$ ($0.9696$) at $N_{grid} = 15$.
If a high precision $>0.99$ is expected, $N_{grid} = 10$ is also a good option with a relative lower accuracy of $0.9112$ and $F_1=0.9469$. 
We next investigate the influence of the thresholds of class probability on the model by fixing the confidence as $0.5$ (see Fig.~\ref{fig11b}). 
The model has good performance at several thresholds of class probability, and to retain the stability of the model, we finally choose the threshold of class probability of $0.5$.     
For the choice of confidence threshold, no significant difference of model's performance is found when the $\kappa \in (0.2, 0.4)$ (see Fig.~\ref{fig11c}).
When $\kappa >0.4$, the $F_1$ and accuracy decrease as the confidence is increased.
Hence a reasonable choice for the confidence threshold is $0.4$.  
The detection model exhibits good performance for MHD wave detection by setting suitable parameters, \ie $N_{grid}$, and thresholds of confidence and class probability.  
In conclusion, the detection model shows promise for MHD wave detection since its simple algorithm uses only 1D input data.   
%

\subsection{\label{sec:3c} Reconstruction of the 2D wave structure}
\begin{figure}[ht] 
\includegraphics[scale=0.7]{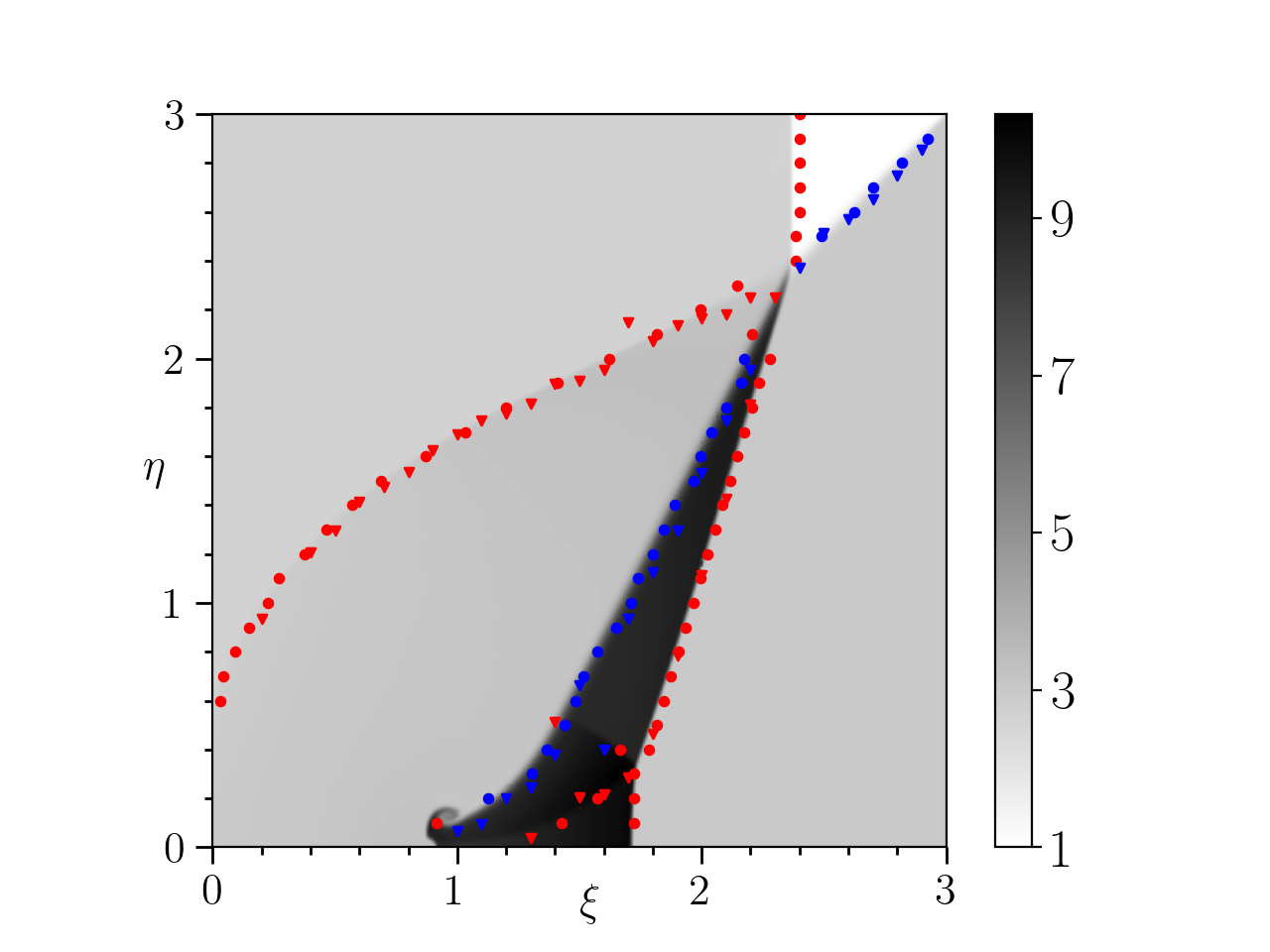}
\caption{\label{fig12} Numerical density field overlaid with detected hydrodynamic shock waves resulting from the detection CNN model. Red and blue points denote hydrodynamic shock and shocked contact, respectively.} 
\end{figure}

\begin{figure}[ht] 
\includegraphics[scale=0.7]{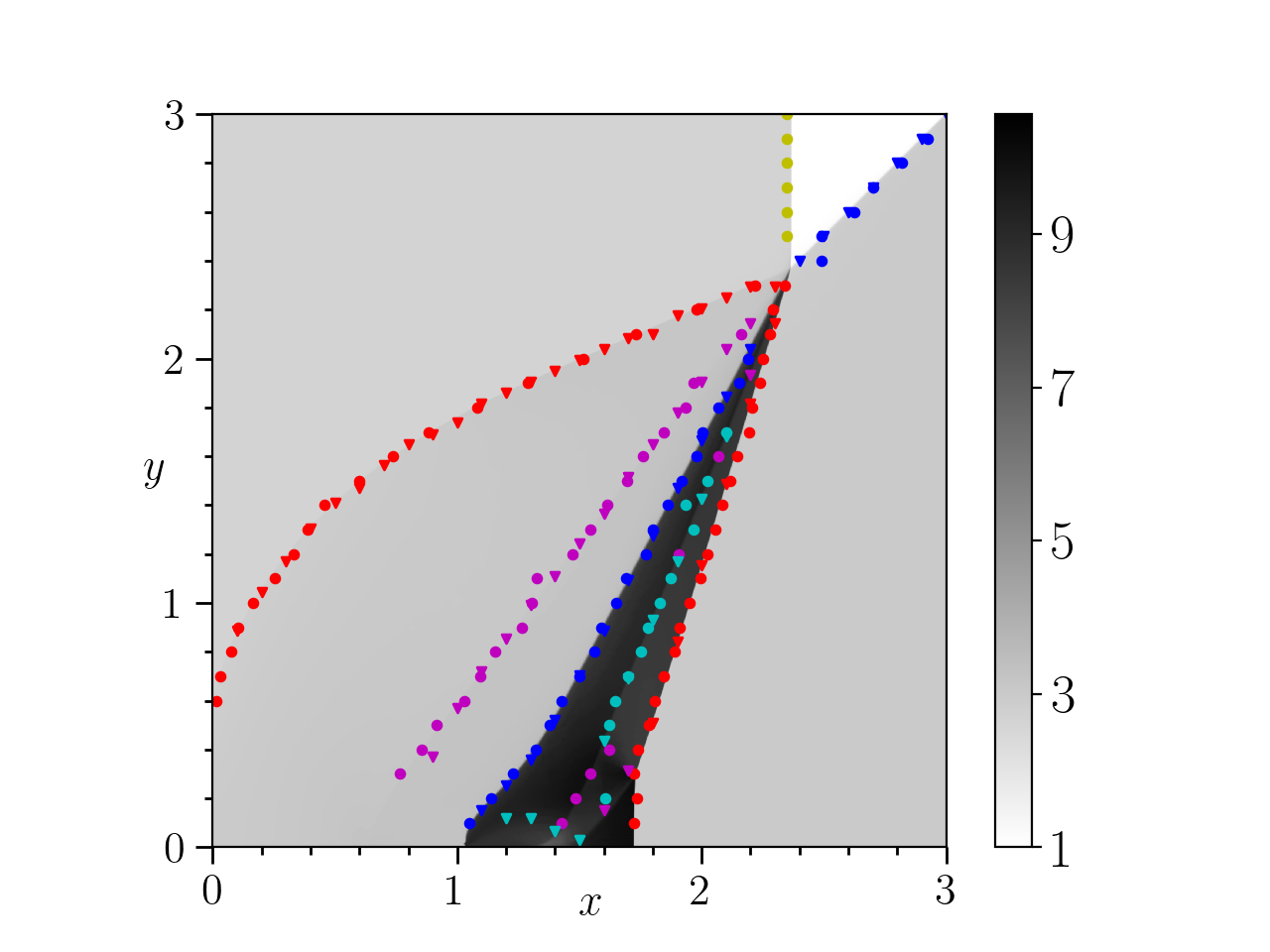} 
\caption{\label{fig13} Numerical density field overlaid with detected MHD shock waves resulting from the detection CNN model. Yellow and blue points denote hydrodynamic shock and shocked contact, respectively; purple, green and red points stand slow shock, $2 \to 4$ intermediate shock and fast shock, respectively.} 
\end{figure}

Finally, we employ the best trained detection model BASED ON 1D input dataset to reconstruct the wave structures in 2D MHD shock refraction. 
The 2D dataset is obtained from numerical simulations using a shock capturing method, and hence all the waves (that are ideally discontinuities), are numerically smeared over a few mesh points. The 2D dataset is split into numerous 1D datasets by taking vertical or horizontal slices of the 2D data.
Each 1D dataset contains the reflected wave $(R)$, shocked contact $(CD)$ and transmitted wave $(T)$ for hydrodynamics, and the $RF, RS, CD, TS$ and $TF$ waves for MHD.
The reconstruction of 2D wave pattern is shown in Fig.~\ref{fig12} for hydrodynamics and Fig.~\ref{fig13} for MHD, respectively. 
For the hydrodynamic case, some waves including shock and shocked contact in the vicinity of triple point are not detected, and also CDs are erroneously detected as shocks.
The non-detection and wrong classification are caused by the waves are in close physical proximity and due to the numerical smearing of these waves due to the underlying numerical method. 
The same reasoning applies for the non-detection and wrong classification in the vicinity of reflection of the waves from the bottom wall. 
However, the simple 1D detection model still successfully reproduces the general structure of hydrodynamic shock refraction. 
In MHD shock refraction, the non-detected waves and wrong classification occur in the vicinity of quintuple point, \ie the point where all the waves $(RF, RS, CD, TS$ and $TF )$ intersect. 
In this case, $RS$ and $TS$ waves are a slow shock and a $I24$ intermediate shock, respectively. Any incorrect classification of the wave types occurs for these two types. 
Also, similar to the hydrodynamics case, the non-detection and wrong classification in the reflection region from the bottom wall is more severe since the wave patterns are  complicated and in physical proximity with each other. 
A small distance away from the triple point in hydrodynamics, and the quintuple point in MHD, the waves are accurately detected. One way further develop a method to connect the dots to reproduce the entire wave front for each wave.
In conclusion, we successfully reproduce the main structure of 2D shock refraction, particularly in vicinity of the point where the all wave interact.   


\newpage
\section{\label{sec:4} Conclusion}
In summary, the present work represents the first instance in the literature of detecting MHD waves in based on a machine learning approach. In MHD shock refraction, there are different wave types which interact with each other in a complex manner. Detecting such waves by traditional methods, such as those to detect hydrodynamic shocks, is cumbersome and error prone. 
The essential idea consists of developing neural networks that employ simple 1D training datasets to detect multi-class MHD waves, such as the intermediate shocks, compound waves and rotational discontinuities, etc. We develop two different models: a fixed output model and a detection model. 
The fixed output model has a high accuracy $>0.99$ with a high recall and precision of more than 0.99. 
The main shortcoming of this model is that the wave number needs to be fixed {\it a priori} for all training datasets. Hence this model has limited applicability for wave detection in general cases.
However, the CNN developed for the output model can be adapted to the second model, \ie the detection model.
For the detection model, we consider the classification and position prediction of waves as only one regression problem. 
The essential idea behind this model is that we divide the input dataset into $N_{grid}$ grid cells. 
A grid cell is responsible to detect a wave if its center is located in that grid cell. 
This algorithm enables the model to be applicable to more general wave detection problem with the accompanying advantage that the number of wave need not be fixed. Each grid predicts the confidence score and class probability which are conditions to decide the detection is correct or not.
The detection model is sensitively affected by the setting of parameters, such as $N_{grid}$, and thresholds of the confidence score and class probability. 
With an appropriately selected set of parameters, the detection model has also a good accuracy of around 0.9 and an efficient $F_1$ score of approximately 0.95. 
One limitation is that the detection model does not exhibit good performance in the vicinity where there are many waves in close physical proximity.  We attribute this to the numerical smearing of these waves, and also the fact that the model is trained with {\em one dimensional} datasets.  
\section*{Acknowledgement}
This research was supported by funding from King Abdullah University of Science and Technology (KAUST) under Grant No. BAS/1/1349-01-01.

\appendix  
\addcontentsline{toc}{section}{Appendices}   

\section{Hyperparameters of the fixed output CCN model for MHD case}
\label{Appa} 
\begin{table}[ht]
\caption{\label{tab1}Hyperparameters of the fixed output CCN model for MHD case}
\begin{ruledtabular}
\begin{tabular}{lll}
\textbf{Hyperparameter} & \textbf{Explanation} & \textbf{Specified or Best value} \\
\itshape $\alpha_{lr}$  &\texttt{Learning rate} & \texttt{1.0 $\times 10^{-3}$}\\
\itshape $n$  &\texttt{Number of features/sample} & \texttt{600}\\
\itshape $k$  &\texttt{Number of output neurons} & \texttt{45}\\
\itshape $n_1$  &\texttt{Number of neurons at $1^{th}$ full connected layer} & \texttt{128}\\
\itshape $n_2$  &\texttt{Number of neurons at $2^{nd}$ full connected layer} & \texttt{64}\\
\itshape $l_{conv}$  &\texttt{Number of 1D convolutional layers} & \texttt{3}\\
\itshape $f_{conv}$  &\texttt{Number of 1D convolutional filters} & \texttt{64}\\
\itshape $s_{conv}$  &\texttt{Size of 1D convolutional filters} & \texttt{5, 4 and 3 in sequence}\\
\itshape $l_{pool}$  &\texttt{Number of 1D max-pooling layers} & \texttt{3}\\
\itshape $s_{pool}$  &\texttt{Size of 1D max-pooling} & \texttt{5}\\
\itshape Optimizer  &\texttt{Stochastic gradient descent} & \texttt{Adam}\\
\itshape $Activation_{full}$   &\texttt{Activation function for full connected layers} & \texttt{Relu}\\
\itshape $Activation_{out}$   &\texttt{Activation function for output layer} & \texttt{Relu $\&$ softmax}\\
\itshape $loss$   &\texttt{Loss function} & \texttt{mse $\&$ categorical-crossentropy }\\
\itshape $metric$   &\texttt{Metric to control the running epoch} & \texttt{rmse $\&$ accuracy}\\
\end{tabular}
\end{ruledtabular}
\footnotetext[1]{Default journal substyle.}
\end{table}
%

\section{Hyperparameters of the detection CCN model for MHD case}
\label{Appb}
\begin{table}[ht]
\caption{\label{tab2}Hyperparameters of the detection CCN model for MHD case}
\begin{ruledtabular}
\begin{tabular}{lll}
\textbf{Hyperparameter} & \textbf{Explanation} & \textbf{Specified value} \\
\itshape $\alpha_{lr}$  &\texttt{Learning rate} & \texttt{1.0 $\times 10^{-3}$}\\
\itshape $n$  &\texttt{Number of features/sample} & \texttt{600}\\
\itshape $N_{grid}$  &\texttt{Number of grids} & \texttt{10}\\
\itshape $Bo$  &\texttt{Number of bounding boxes} & \texttt{1}\\
\itshape $C$  &\texttt{Number of classes} & \texttt{8}\\
\itshape $k$  &\texttt{Number of output neurons} & \texttt{$N_{grid} \times (3Bo+C)$}\\
\itshape $n_1$  &\texttt{Number of neurons at $1^{th}$ full connected layer} & \texttt{256}\\
\itshape $n_2$  &\texttt{Number of neurons at $2^{nd}$ full connected layer} & \texttt{128}\\
\itshape $l_{conv}$  &\texttt{Number of 1D convolutional layers} & \texttt{3}\\
\itshape $f_{conv}$  &\texttt{Number of 1D convolutional filters} & \texttt{64}\\
\itshape $s_{conv}$  &\texttt{Size of 1D convolutional filters} & \texttt{5, 4 and 3 in sequence}\\
\itshape $l_{pool}$  &\texttt{Number of 1D max-pooling layers} & \texttt{3}\\
\itshape $s_{pool}$  &\texttt{Size of 1D max-pooling} & \texttt{5}\\
\itshape Optimizer  &\texttt{Stochastic gradient descent} & \texttt{Adam}\\
\itshape $Activation_{full}$   &\texttt{Activation function for full connected layers} & \texttt{Relu}\\
\itshape $Activation_{out}$   &\texttt{Activation function for output layer} & \texttt{Relu}\\
\itshape $loss$   &\texttt{Loss function} & \texttt{mse}\\
\itshape $metric$   &\texttt{Metric to control the running epoch} & \texttt{accuracy}\\
\end{tabular}
\end{ruledtabular}
\footnotetext[1]{Default journal substyle.}
\end{table}
%
\newpage
\section*{Reference}
\bibliography{apssamp}
\end{document}